\documentclass[aps,prx,twocolumn,floats,epsfig,pdflatex]{revtex4-2}
\usepackage{amsmath}
\usepackage{amsfonts}
\usepackage{graphicx}

\usepackage{amssymb}
\usepackage{amsbsy}
\usepackage{color}
\usepackage{ulem}

\usepackage{hyperref}
\hypersetup{
	colorlinks=true,
	citecolor=blue,
	linkcolor=blue,
	urlcolor=blue}

\newcommand{\bra}[1]{\langle{#1}|}
\newcommand{\ket}[1]{|{#1}\rangle}

\newcommand{\beq}{\begin{equation}}
\newcommand{\eeq}{\end{equation}}
\newcommand{\bea}{\begin{eqnarray}}
\newcommand{\eea}{\end{eqnarray}}

\begin{document}

\title{Emergence of a quasi-ergodic steady state in a dissipative Tavis-Cummings array}
\author{Debabrata Mondal$^1$, K. Sengupta$^2$, and S. Sinha$^1$}
\affiliation{$^1$ Indian Institute of Science Education and Research-Kolkata, Mohanpur, Nadia-741246, India.\\
$^2$School of Physical Sciences, Indian Association for the
Cultivation of Science, Kolkata 700032, India}
\date{\today}

\begin{abstract}
In an atom-photon interacting system described by the Tavis-Cummings Hubbard (TCH) model, we demonstrate the emergence of a quasi-steady state in a dissipative environment that exhibits intriguing ergodic behavior. The TCH model undergoes a dissipative transition from normal to superradiant phase hosting a gapped Higgs and gapless Goldstone modes. However, in a large region of the phase diagram, the instability of the Goldstone mode leads to the disappearance of the stable superradiant phase. In this regime, the decorrelator dynamics reveals light cone spreading of the perturbations and positive Lyapunov exponent, indicating enhanced fluctuations. Remarkably, a quasi-steady state emerges under quench dynamics in this unstable regime; in this state, a class of collective quantities such as site averaged photon number and atomic excitations approach a steady value, in spite of the large temporal fluctuations in corresponding microscopic variables. This quasi-steady state describes an incoherent fluid of photons with significant phase fluctuation. The phase space dynamics reveals a fascinating ergodic behavior in the presence of dissipation leading to the characterization of the dynamical variables into two distinct classes. The first class includes site-averaged photon numbers and atomic excitations; these exhibit a stationary distribution regardless of the initial condition indicating ergodic behavior. The second class of variables, particularly those related to photon or atom phases, in contrast, retain information about the initial conditions, resulting in a violation of ergodicity for finite size systems. Additionally, the dynamical variables of the ergodic class exhibit a fascinating {\it collective scarring} phenomenon as the peak of their distribution is attracted towards the unstable steady state, analogous to the single particle quantum scar. We discuss the relevance of our findings in the current experiments.

\end{abstract}


\maketitle

\section{Introduction}

Emulation of strongly correlated quantum systems using ultracold
atoms coupled to suitably tuned coherent laser sources have seen
tremendous theoretical and experimental progress in recent years
\cite{rev1,rev2,rev3,rev4,rev5,rev6,rev7,Helmut_Ritsch,AJ_Daley,rev7,rev8,rev9,rev10,rev11,rev12,rev13,rev14}. These systems allow one to study the ground states of the correlated models in parameter regimes that are normally not attainable in standard
condensed matter setups. Thus they can be used for exotic  symmetry
broken quantum ground states and their corresponding phase
transitions \cite{exp1,exp2,exp3,exp4,exp5,exp6, th1,th2,th3,th4,th5,th6,th7,th8,th9,  ryd1,ryd2,ryd3,ryd4,ryd5,ryd6,ryd7}; moreover, they also serve as ideal platforms for analyzing the quantum dynamics of these models
\cite{rev6, rev7,rev8}. The latter set of studies has provided a wealth
of information regarding long-time dynamics of strongly correlated
non-integrable quantum models \cite{rev6,rev8,rev9,rev10,rev11,rev12,rev13,rev14}; thus these
atom-photon systems serve as experimental test bed for several
important tenets in the field of non-equilibrium phenomena
such as the eigenstate thermalization hypothesis (ETH).

A separate class of experimental setups involving atomic cavities coupled to photons have also been put forth as emulators of the
strongly interacting bosonic models
\cite{jcrefs0,jcrefs1,jcrefs2,jcrefs3,tc1,tc2,AM_Rey_1,AM_Rey_2,AJ_Daley_2}.
Such cavity systems typically have $N$ two-level oscillators,
serving as model of atoms, that are usually coupled to a single photonic cavity mode. These atom-photon interacting systems can be described by the well-known
Jaynes-Cummings model for a single atom in the cavity ($N=1$) \cite{jcrefs1,jcrefs2,jcrefs3}, which has also been extended for $N>1$ and the extended model is referred to as the Tavis-Cummings (TC) model \cite{tc1,tc2}. Both these models have been extensively studied in the literature in various contexts. Notably, the TC model undergoes a superradiant phase transition for
$N \rightarrow \infty$ \cite{sp1,sp2,sp3,sp4,sp5}.

Furthermore, many body physics with photons can be explored in the lattice models of cavity arrays, where each cavity serves as the
site of a $d-$dimensional hypercubic lattice. These individual
sites are coupled with their nearest neighbors via a photon mode \cite{jcrefs0,Angelakis}.
The experimental realization
of such systems have long been achieved using cavity arrays \cite{JCH_Plenio,Coupled_cavities_1,Coupled_cavities_2} and
multimode cavities in circuit QED setups \cite{cavref1,cavref2,cavref3,cavref4,cavref5,cavref6}; they have also been realized in ion-trap systems \cite{itref1}.
Such experimental systems offer more control over individual cavities
in the array; this makes it easier to probe site-resolved local
dynamics in such setups compared to their ultracold atom
counterparts.

These arrays of coupled cavities are known to host superfluid-insulator transitions \cite{jcref4,jcref5,jcref6,jcrefs4} and supersolid phases \cite{jcref7}.
Particularly, the Tavis-Cummings lattice model has been studied in
the quantum regime using density matrix renormalization group
technique and variational cluster
methods \cite{knapp1,baksic1,pelster1,fazio1,nori1,fan1}. These
studies have revealed the presence of ground states with fixed
photon numbers mimicking the Mott insulating states of the
Bose-Hubbard model. In addition, the
model supports a superradiant (SR) phase with variable photon number,
which corresponds to the superfluid phase of the Bose-Hubbard model
\cite{knapp1}. The quantum phase transition between these phases has
also been studied \cite{fazio1,knapp1,nori1,fan1}; they belong to
the same universality class as the superfluid-insulator transition
of the Bose-Hubbard model.

The atom-photon systems realized in these setups are typically
dissipative in nature due to photon loss to the environment and
decay of the oscillator degrees of freedom via incoherent radiation
to the non-cavity modes \cite{pelster1}. Moreover, in a typical
experimental realization, the cavities are pumped with incoherent
radiation leading to the generation of excitations of the cavity modes.
Theoretically, the treatment of such pumped and dissipative system
necessitates the use of density matrices \cite{dm1}. The equation of
motion for these density matrices encodes the effect of dissipation
and pumping through a Lindblad operator; the form of such operators
in the present setting is well-known \cite{pelster1,opref1}. Such
systems have been analyzed for $N=2$; however, their large $N$ limit,
which allows a systematic controlled semiclassical description of
the dynamics of the system has not been explored. The
non-equilibrium quench dynamics of such a dissipative and pumped TC
model, which provides information about steady states accessed by such a
system, has also not been studied so far.

The steady states of such dissipative quantum systems typically describe non-equilibrium phases. In contrast, the steady states in closed quantum systems are known to be thermal and are related to the ergodic behavior of quantum states; their behavior can be understood using ETH. Such ergodic behavior has been studied extensively in recent years \cite{Ueda,Bloch_1,Schmiedmayer,Greiner,Weidem,M_Rigol_1,Hauke_1,rev6,rev7,rev9,L_Santos_1}.
The recent observation of many-body quantum scars in experiments involving an array of Rydberg atoms and in circuit quantum electrodynamics (QED) setups has emerged
as a source of weak ergodicity violation in closed systems displaying intriguing athermal behavior, which has become a topic of intense research \cite{exp6,Scar_superconducting_qubit}. However, the fate of such thermal steady states and the issue of ergodicity in the presence of dissipation is a relatively unexplored area. In this context, attempts have been made to diagnose chaos in the dissipative quantum many body systems with non-unitary dynamics \cite{Haake_dissipation,Prosen_1,Prosen_2,Prosen_3,int_chaos,Manas_1,SYK_chaos, Minganti,Prosen_4}. Since dissipation is inherent in most experimentally realizable systems, it becomes a pertinent concern to explore scarring phenomena in dissipative systems.

The central question we ask in this work is the following. Can there be a manifestation of ergodicity in the presence of dissipation and if so what is the nature of such dissipative steady states? We address these issues concerning ergodicity and the signature of unstable dynamical states, akin to scarring phenomena, in this atom-photon interacting dissipative many body systems described by the TCH model in the semiclassical (large $N$) limit. We note in this context that such collective models with suitable semiclassical limit have become ideal platforms for exploring the classical route to ergodicity and its deviation \cite{Haake_2012,Altland_Haake_2012, D_mondal_scar,KCT}. We present a synopsis of our main results in Sec.\ \ref{synop}.

The plan for the rest of the paper is as follows. In Sec.\ \ref{modeltc} we define the model and discuss some of its features. This is followed, in Sec.\ \ref{sstate}, by a discussion of the homogeneous steady states and their stability. Next, in Sec.\ \ref{dissdyn}, we discuss the quench dynamics in the unstable regime. The behavior of the decorrelators is discussed in Sec.\ \ref{deco}, the nature of the collective variables is charted out in Sec.\ \ref{steady}, and the collective scarring phenomenon is discussed in Sec.\ \ref{scar1}. Finally, we discuss our main results, chart out experiments which can verify them, and conclude in Sec.\ \ref{conc}. The phases of the model without dissipation and pumping and some details of instability of the collective excitations around the steady states are detailed in the appendices.

\subsection{Synopsis of main results}
\label{synop}

In this work, we study Tavis-Cummings Hubbard (TCH) model \cite{knapp1,fazio1} which describes an array of cavities, each containing N atoms that interact with a single photon mode. The coupling between these cavities results in nearest neighbor photon hopping \cite{JCH_Plenio,npj_hopping}. This system is subjected to dissipation arising from natural photon loss from cavities as well as atom pumping \cite{Dissipation3,Dissipation4,Dissipation5,Dissipation6,pelster1,Sayak_ADM}. We study the model in the large $N$ limit; this limit allows us to study the model semiclassically to analyze the dynamics and steady states. The main results obtained from our study are as follows:

\begin{itemize}

\item
From the homogeneous steady states of this system, we obtain normal phases with fixed photon number (${\rm NP}_0$, ${\rm NP}_1$) as well as the superradiant (SR) phases with variable photon number. We also obtain the low-lying excitations of these phases such as the gapless Goldstone and the gapped Higgs modes in the SR phase. In the presence of dissipation, the nature of the transition from normal to SR phase changes along the commensurate line; this is reflected in the dispersion of Higgs excitation. For a wide parameter regime, we find that the instability of the low-lying Goldstone mode leads to the destabilization of these phases.

\item
We demarcate the unstable regime from light-cone like spreading of fluctuations during dynamics of decorrelators; such spreading also indicates chaos exhibiting a positive Lyapunov exponent. Interestingly, the saturation value of the decorrelator is found to be linked to an order associated with a unique quasi-steady state that emerges in the unstable regime.

\item
Using quench dynamics from arbitrary initial state, we identify the emergence of a unique quasi-steady state as indicated above; this state describes an inhomogeneous fluid of incoherent photons with enhanced phase fluctuations. In this state, the collective site-averaged variables, such as total photon number $n =\sum_i n_i$ and total atomic excitation $s_z= \sum_i s_{iz}$ reach a steady state value, while the microscopic on-site variables $n_i$ and $s_{iz}$, exhibit large temporal fluctuations.
However, the distributions of these variables become stationary, which determine the statistical properties of the fluid. On the contrary, the collective quantities involving phase variables, such as the atomic superfluid order parameter do not approach to a steady value.

\item
Our study therefore unveils an intriguing ergodic behavior of the quasi-steady state, which indicates the presence of two distinct classes of collective variables in these systems. The first class of such collective variables, which include $n$ and $s_z$, attain a steady state value {\it irrespective of the initial state}. This loss of memory of the initial condition indicates  ergodic behavior in spite of the presence of dissipation. The steady state value reached can be described, in the parlance of dynamical systems, as a fixed point in the phase space of these collective variables. Furthermore, at the steady state, the site distribution of these variables for a single initial condition matches with the corresponding distribution at any arbitrary site for an ensemble of random initial states, which is the key feature of ergodicity.

In contrast, the second class of collective variables related to the phase degrees of freedom, which includes the superfluid order parameter of the atom modes, $\mathcal{M}_s$, in the SR phase, retains the memory of initial conditions and forms a distribution for an ensemble of randomly chosen initial states. However, the most probable values of these order parameters remain vanishingly small. 
In the parlance of dynamical systems, these variables form a fixed surface. Thus the quasi-steady state displays a unique dichotomy wherein one class of collective variable
displays ergodic behavior leading to a fixed point in phase space whereas the other class shows a clear violation of ergodicity yielding a fixed surface. To the best of our knowledge, such a phenomenon has not been reported in the context of dissipative atom-photon systems
so far.

\item
Notably, we identify the influence of unstable fixed points on the formation of this quasi-steady state, which we refer to as {\it collective scarring} in analogy with quantum scars observed in chaotic systems. The distributions of the ergodic class of variables $\{n,s_z\}$ are peaked around the unstable homogeneous fixed points; thus the most-probable
value of such microscopic variables are attracted to the unstable homogeneous steady states.
This leads to higher phase space density around the unstable fixed point and is therefore reminiscent of scarring in classical systems. Furthermore, a crossover from one unstable steady state to another takes place, due to which the quasi-steady state smoothly interpolates between the stable SR and ${\rm NP}_0$ phases.

\end{itemize}

The present study therefore not only predicts a unique phase of incoherent photonic fluid in experimentally realizable coupled atom-photon interacting systems, but also poses new questions regarding ergodicity properties of such systems in the presence of dissipation.

\section{The model}
\label{modeltc}

We consider an array of cavities arranged on a $d$-dimensional
hypercubic lattice, where each cavity contains $N$ two level atoms
with energy gap $\hbar\omega_0$. These atoms interact with a single
mode cavity field of frequency $\omega$. The interaction between atoms and photons within each cavity can be described by the TC model \cite{tc1}, given by,
\begin{eqnarray}
\hat{\mathcal{H}}_{\rm TC}^{(i)} = \omega\,
\hat{a}^{\dagger}_i\hat{a}_i+\omega_0\,
\hat{S}_{iz}+\frac{\lambda}{\sqrt{2S}}(\hat{a}_i\,\hat{S}_{i+}+\hat{a}^{\dagger}_i\,\hat{S}_{i-}).
\label{TCM}
\end{eqnarray}
Here $i$ denotes the site index of the cavity,
$\hat{a}_i$($\hat{a}_i^{\dagger}$) is the annihilation (creation) operator for the
photon mode coupled to the atoms in the cavity, and the collective spin
operators $\hat{S}_{i\pm},\hat{S}_{iz}$ of magnitude $S=N/2$ are associated with the $N$ two level systems at the corresponding site.
%
The interaction strength between the atoms and
electromagnetic mode is $\lambda$. Here and in the rest of this
work, we set $\hbar=1$ and the detuning to be positive $(\omega-\omega_0)>0$.
\begin{figure}
\centering
\includegraphics[width=\columnwidth]{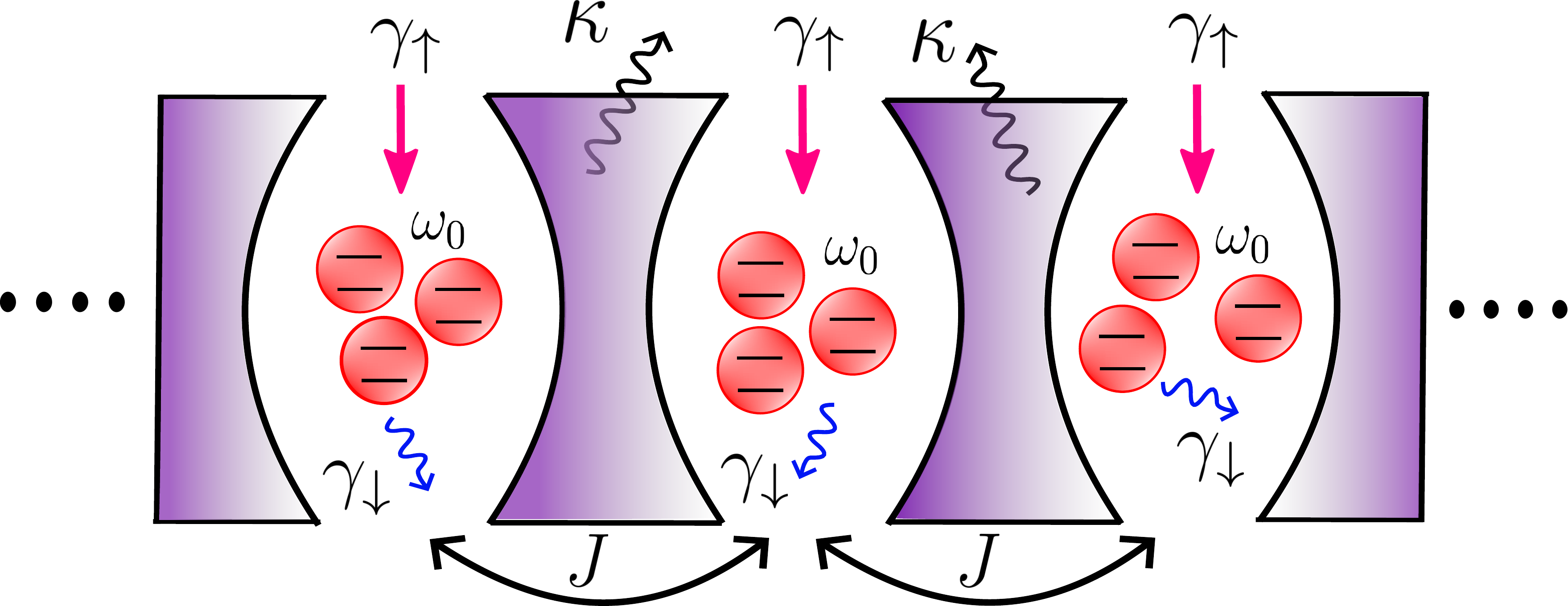}
\caption{Schematic of the dissipative Tavis-Cummings lattice, consisting of an array of optical cavities arranged in one-dimension (1D). Each cavity has a single mode photon field of frequency $\omega$. The nearest-neighbor lattice sites are coupled due
to photon hopping of amplitude $J$. Each cavity interacts with a collection of two-level atoms each having an energy gap $\hbar \omega_0$. The cavities lose photons with rate $\kappa$. $\gamma_{\downarrow}$ represents the rate of collective decay of the atoms. In addition, the atoms are incoherently pumped with rate $\gamma_{\uparrow}$.}
\label{fig1}
\end{figure}
The Hamiltonian describing an array of such cavities can be written as,
\begin{eqnarray}
\hat{\mathcal{H}} &=& -J\sum_{\langle
ij\rangle}(\hat{a}_i^{\dagger}\hat{a}_j+\hat{a}_j^{\dagger}\hat{a}_i)+\sum_i\hat{\mathcal{H}}_{\rm
TC}^{(i)} \label{Hamiltonian},
\end{eqnarray}
where the first term represents the nearest neighbor hopping of
photons with amplitude $J$, $\langle ij\rangle$ implies that the
indices $i$ and $j$ correspond to the nearest neighbor cavities on the
hypercubic lattice with the coordination number $z_0=2d$.  We denote this model as the Tavis-Cummings Hubbard (TCH) model. Note that the Hamiltonian $\hat{{\mathcal H}}$ exhibits ${\rm U}(1)$ symmetry, leading to the conservation of the total excitation,
\begin{eqnarray}
\hat{\mathcal{N}} &=& \sum_i\hat{\mathcal{N}_i} = \sum_i
(\hat{a}_i^{\dagger}\hat{a}_i+S+\hat{S}_{iz}). \label{numcon}
\end{eqnarray}
Here, $S+\hat{S}_{iz}$ describes the atomic excitations in each cavity.
In the absence of dissipation, this model can be described by the Hamiltonian $\hat{\mathcal{H}}-\mu \hat{\mathcal{N}}$ in the grand canonical ensemble, where $\mu$ is the chemical potential corresponding to the total excitation number \cite{knapp1,fazio1}.

In realistic situations, such systems of coupled cavities experience dissipation arising from photon loss and are usually subjected to incoherent	pumping.
In the presence of these processes, it is customary to analyze the evolution of the density matrix $\hat{\rho}$ of the system, which is governed by the master equation of the Lindblad form \cite{Breuer},
%
\begin{eqnarray}
\dot{\hat{\rho}} &=& -i[\mathcal{\hat{H}},\hat{\rho}]+ \kappa \sum_i\mathcal{L}[\hat{a}_i] \nonumber\\
&& +\frac{1}{S}\sum_i(\gamma_{\uparrow} \mathcal{L}[\hat{S}_{i+}]+
\gamma_{\downarrow}
\mathcal{L}[\hat{S}_{i-}]),
\label{dmeq1}
\end{eqnarray}
with $\mathcal{L}[\hat{\mathcal{O}}] = \frac{1}{2} \left(2\hat{\mathcal{O}}\rho \hat{\mathcal{O}}^{\dagger}-\hat{\mathcal{O}}^{\dagger}\hat{\mathcal{O}}\rho-\rho \hat{\mathcal{O}}^{\dagger}\hat{\mathcal{O}}\right)$ describing the dissipative process corresponding to the operator $\hat{\mathcal{O}}$.
The photon loss characterized by $\mathcal{L}[\hat{a}_i]$ occurs
due to the partial reflectivity of a cavity and leads to the incoherent
dynamics with a decay rate $\kappa$. The system also
experiences the collective decay of the spins $\mathcal{L}[\hat{S}_{i-}]$ with rate $\gamma_{\downarrow}$ due to radiation into non-cavity modes \cite{Sayak_ADM,MF_1}. 
In addition, $\mathcal{L}[\hat{S}_{i+}]$ represents the incoherent pumping of atoms to the excited state at a rate $\gamma_{\uparrow}$.
The steady state with nonvanishing photons can emerge due to the competition between  these decay processes and incoherent pumping when $\gamma_{\uparrow}>\gamma_{\downarrow}$, as the latter process results in creating additional photons in the system. Such pumping of atoms can be achieved by  engineering a nonequilibrium heat bath with negative temperature, which can be modeled by a collection of  inverted harmonic oscillators, as discussed in \cite{dm1,Spin_pumping_1}. Similar to Lasers, the atomic excitation can be done optically, which can be effectively be described as an incoherent pumping process \cite{Spin_pumping_2}.
%
Notably, in presence of these collective processes, the total spin at each site remains conserved. The schematic of the dissipative TCH model is illustrated in Fig.\ref{fig1}.

The dynamics of statistical average $\langle \hat{A}\rangle = {\rm Tr}(\hat{A}\hat{\rho})$ of any operator $\hat{A}$ can be described by
$d\langle\hat{A}\rangle/dt = {\rm Tr}(\hat{A}\dot{\hat{\rho}})$. In what
follows, we investigate the state of this dissipative TCH model, where photon loss can be balanced by collective pumping in the large $N$ limit, using the semiclassical equations of motion (EOM) for the observables.

The scaled observables $(\hat{x}_{i} + i\hat{p}_{i})/\sqrt{2} = \hat{a}_{i}/\sqrt{S}$ and $\vec{\hat{s}}_{i} = \vec{\hat{S}}_{i}/S$ satisfy the commutation relations $[\hat{x}_i,\hat{p}_i]=i/S$ and $[\hat{s}_{i l},\hat{s}_{i m}]= i\epsilon_{lmn}\hat{s}_{i n}/S$, for which $1/S$ (or 2/$N$) plays the role of effective Planck constant.
Consequently, for $S\gg 1$ (equivalently $N\gg1$), the scaled variables effectively become classical.
Due to such classical correspondence in large $N$ limit, the pure state of this system, in absence of dissipation can be described semiclassically by the variational wave function \cite{Coherent_state},
\begin{eqnarray}
\ket{\Psi_c} &=&  \prod_i\ket{\alpha_i}\otimes
\ket{\mathcal{Z}_i},
\label{varwav1}
\end{eqnarray}
where $\ket{\mathcal{Z}_i}$ and $\ket{\alpha_i}$ are the coherent
states of spin with $\mathcal{Z}_i = \tan(\theta_i/2)\exp(\iota \phi_i)$ and bosonic modes respectively at site $i$. 
The conjugate spin variables at each site $s_{iz} = \cos\theta_i$ and
$\phi_i$ represent the orientation of a classical spin vector
\begin{eqnarray}
\vec{S}_i=S(\sin{\theta_i}\cos{\phi_i},\sin{\theta_i}\sin{\phi_i},\cos{\theta_i})
\label{spinparam}
\end{eqnarray}
on the Bloch sphere. Here $\alpha_i/\sqrt{S} = (x_i+ ip_i)/\sqrt{2} = \sqrt{n_i}\exp(-i\psi_i)$,
where $n_i$ denotes the photon number on the $i^{\rm th}$ site and
$\psi_i$ indicates its phase.
%
%
Therefore, such coherent states can be used to treat the above operators classically, since their fluctuations relative to the mean vanish in the limit $S\rightarrow\infty$. As a result, these coherent states are better suited for representing a classical phase space point. 

Using the wave function given in Eq.\eqref{varwav1}, in the semiclassical limit, the scaled Hamiltonian in the grand canonical ensemble $\mathcal{H}_c=\langle \hat{\mathcal{H}}-\mu\hat{\mathcal{N}}\rangle/S$ and the excitation number $\mathcal{N}_c=\langle \hat{\mathcal{N}}\rangle/N$ can be written in terms of the conjugate variables  $\{\psi_i,n_i,\phi_i, s_{iz}\}$ as,
\begin{subequations}
\begin{align}
	\mathcal{H}_{c} &= -2J\sum_{\langle ij\rangle}\sqrt{n_in_j}\cos(\psi_i-\psi_j)+\sum_i\bigg[(\omega-\mu)\,n_i\bigg.\nonumber\\
	 &+\left.(\omega_0-\mu)\,s_{iz}+
	\lambda\sqrt{2n_i}\sqrt{1-s_{iz}^2}\,\cos(\phi_i-\psi_i)\right],\label{hameq}\\
	\mathcal{N}_{c}&= \sum_i\left(\frac{n_i+1+s_{iz}}{2}\right).
	\label{excitationeq}
\end{align}	
\end{subequations}
Here, the classical Hamiltonian $\mathcal{H}_{c}$ and the total excitation number $\mathcal{N}_{c}$ are scaled by the spin magnitude $S$ and the number of atoms $N$, respectively.
The EOM of the above mentioned classical variables can be obtained from the classical Hamiltonian $\mathcal{H}_c$ (see Appendix.\ref{appa}), which can alternatively be derived from the Heisenberg equations of motion of the corresponding operators.

Next, we focus on the dissipative model, where the dynamics of the
statistical average of any operator $\langle \hat{A}\rangle = {\rm Tr}(\hat{A}\hat{\rho})$ can be calculated from the master equation Eq.\eqref{dmeq1} of the density matrix $\hat{\rho}(t)$, using $d\langle\hat{A}\rangle/dt = {\rm Tr}(\hat{A}\dot{\hat{\rho}})$. Similar to the semiclassical wavefunction, such a large $N$ system allows us to approximate the total density matrix in the product form,
\begin{equation}
\hat{\rho} = \prod_i \hat{\rho}_{i,s}\otimes \hat{\rho}_{i,p}
\label{var_dm}
\end{equation}
where $\hat{\rho}_{i,s(p)}$ represents the density matrix corresponding to the spin (photon) degrees of freedom at site $i$.

This product density matrix enables us to factorize the average
of product operators $\langle \hat{A}\hat{B}\rangle \approx \langle \hat{A}\rangle \langle\hat{B}\rangle$, that is equivalent to the mean-field approximation, which is a standard technique for such systems with large $N$ \cite{Sayak_ADM,MF_1,Dissipation5,Dissipation6}. 
For $N\gg 1$, the average value of the above mentioned scaled observables correspond to the classical variables, as described by the coherent state, given in Eq.\eqref{varwav1} \cite{MF_1}.
Within the above-mentioned mean field approximation, we obtain the following semiclassical EOM using the Lindblad master equation Eq.\eqref{dmeq1} in the large $N$ limit,
\begin{subequations}
    \begin{eqnarray}
        \dot{n}_i &=& -\lambda\sqrt{2n_i}\sqrt{1-s^2_{iz}}\sin(\phi_i-\psi_i)-\kappa n_i \label{EOM_n}\\
        &-&2J\sum_{\delta}\sqrt{n_in_{i+\delta}}\sin(\psi_i-\psi_{i+\delta})\nonumber\\
        \dot{s}_{iz} &=& \lambda\sqrt{2n_i}\sqrt{1-s_{iz}^2}\,\sin(\phi_i-\psi_i)+f_c (1-s_{iz}^2)\qquad\label{EOM_sz}\\
        \dot{\psi}_i &=& \omega+\frac{\lambda\sqrt{1-s_{iz}^2}}{\sqrt{2n_i}}\cos(\phi_i-\psi_i)\nonumber\\
        &-&J\sum_{\delta}\sqrt{\frac{n_{i+\delta}}{n_{i}}}\cos(\psi_i-\psi_{i+\delta})\\
        \dot{\phi}_i &=&
        \omega_0-\frac{\lambda\sqrt{2n_i}\,s_{iz}}{\sqrt{1-s_{iz}^2}}\cos(\phi_i-\psi_i),
    \end{eqnarray}
    \label{EOM_dissipation}
\end{subequations}
where the net incoherent pumping strength is characterized by $f_c = \gamma_{\uparrow}-\gamma_{\downarrow}$. 
The exactness of the mean field equations for such collective systems in the limit $S\rightarrow\infty$ has been shown in \cite{Mean_field_1,Mean_field_2}.
However, the beyond mean field effects such as quantum fluctuations can be investigated by systematically incorporating the evolution of the cumulants \cite{Mean_field_1}.
Next, we analyze Eq.\eqref{EOM_dissipation} to explore the dynamics of the dissipative TCH model both analytically and numerically. Throughout the work, we consider $\hbar,k_{\rm B}=1$ and $\omega = 2,\omega_0 = 0.5,f_c=0.1$ for all numerical results. We also scale the energy (time) by $Jz_0$ (1/$Jz_0$), which is equivalent to considering $Jz_0 = 1$.

\section{Steady states and excitation spectra around them}
\label{sstate}

In this section, we study the homogeneous phases of the dissipative model obtained from the above EOM (Eq.\eqref{EOM_dissipation}) by considering $n_i=n^*$, $s_{iz}=s_z^*$, and $\phi_i=\phi^*$. The relative phase of the photon between the consecutive sites $\psi_i-\psi_{i+\delta}$ can be either $0$ or $\pi$, corresponding to the steady states. In this work, we only focus on the case $\psi_i=\psi_{i+\delta}=\psi^*$, since we find that the other case with $\pi$ phase difference does not play any significant role in the dynamics of the system.

The results obtained from such an analysis are as follows. First, we obtain a class of steady states where the photon number $n_i= n^{\ast}=0$; there are two such phases corresponding to $s_z^{\ast}=\pm 1$. The state corresponding to $s_z^{\ast}=-1$ is termed as NP$_0$ while that with $s_z^{\ast}=1$ is denoted by NP$_1$. The excitation number corresponding to the normal phase NP$_0$ (NP$_1$) is $\mathcal{N}_c=0\,(1)$.
In addition, we also find superradiant (SR) phases with non-zero photon number
$n^*$, which varies with system parameters. We note that all of these states exist as the ground states of $\mathcal{H}_c$ (Eq.\eqref{hameq}) in the absence of pumping and dissipation, as shown in Appendix.\ref{appa}.
\begin{figure}
\includegraphics[width=\linewidth]{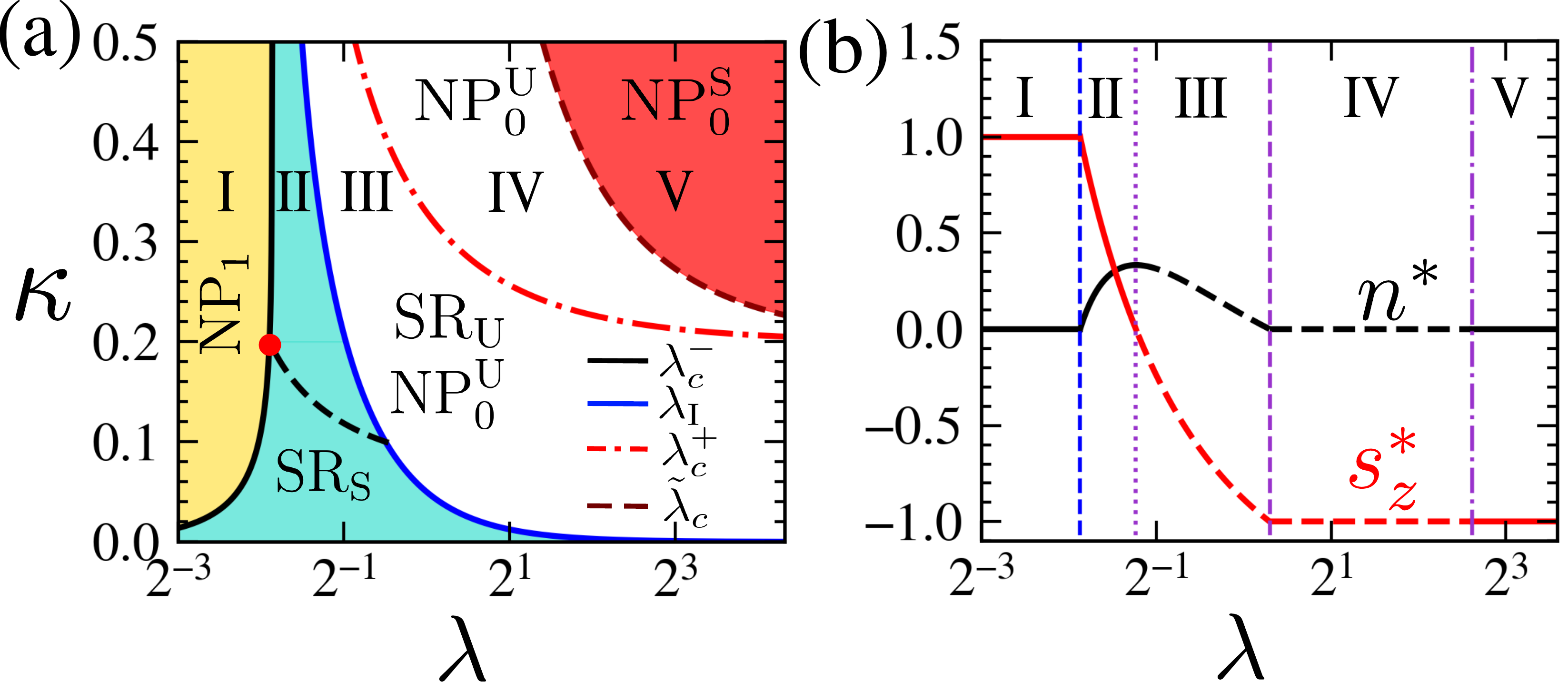}
\caption{{\it Steady states of the dissipative TCH model and their stability regimes:} (a) The steady state phase diagram in $\kappa-\lambda$ plane. The stable (unstable) superradiant and normal phases are denoted by SR$_{\rm S}$ (SR$_{\rm U}$) and NP$_0^{\rm S}$ (NP$_0^{\rm U}$) respectively. The boundary between the normal phase NP$_1$ and SR$_{\rm S}$ is indicated by the solid black line. The superradiant phase remains stable in the light blue colored regime and the unstable SR$_{\rm U}$ meets the unstable normal phase
NP$_0^{\rm U}$ on the dashed dotted red line. The normal phase becomes stable (NP$_0^{\rm S}$) in the red colored regime corresponding to $\lambda>\tilde{\lambda}_c(\kappa)$. The line of commensurate excitation number is denoted by the black dashed line. The red dot corresponds to the critical point ($\lambda^*_{f_c},\kappa^*$). (b) The variation of photon number $n^*$ and atomic population $s_z^*$ of the steady states with coupling strength $\lambda$ corresponding to the decay rate $\kappa = 0.4$. The stable (unstable) steady states are marked by solid (dashed) lines.  All energies (time) are measured in the unit of hopping amplitude $J(1/J)$. In this and the rest of the figures, we consider $\hbar, k_{\rm B} = 1$ and $\omega=2,\omega_0=0.5, Jz_0=1.0, f_c=0.1$. }
\label{fig2}
\end{figure}

It is important to mention that the transformations $\phi_i\rightarrow
\phi_i+\theta$ and $\psi_i\rightarrow \psi_i+\theta$ leave the EOM (Eq.\eqref{EOM_dissipation}) unchanged, indicating the $U(1)$ symmetry of the system even in the presence of these incoherent decay and pumping process, however, the total excitation number no longer remains a conserved quantity. This $U(1)$ symmetry allows for a superradiant phase with a nonzero photon number. However, to obtain this phase as a steady state, we need to switch to a rotating frame with the frequency $\mu_0$,
which can be done by replacing $\phi_i\rightarrow\phi_i+\mu_0 t$ and
$\psi_i\rightarrow\psi_i+\mu_0 t$ in Eq.\eqref{EOM_dissipation}
\cite{pelster1}. This is equivalent to using the free energy to obtain the EOM in the non-dissipative case, where $\mu_0$ plays the role of chemical potential. A straightforward analysis indicates that the frequency $\mu_0$ of the rotating frame for the SR phase is
\begin{eqnarray}
\mu_0 &=&
\omega-Jz_0-\frac{1}{2}\sqrt{\frac{\kappa}{f_c}}\,\sqrt{2\lambda^2-\kappa
f_c}
\end{eqnarray}
where $z_0=2d$ is the coordination number of the lattice. In this frame, the SR states are characterized by,
\begin{eqnarray}
s_{z}^{\ast} &=&
-\frac{\kappa}{2f_c}+\sqrt{\frac{\kappa}{f_c}}\left(\frac{\omega-\omega_0-
Jz_0}{\sqrt{2\lambda^2-f_c\kappa}}\right)\nonumber\\
n^{\ast} &=&
\frac{f_c}{\kappa}(1-s_{z}^{*2}),\,\,\,\sin(\phi^*-\psi^*) =
-\frac{\sqrt{f_c\kappa}}{\lambda \sqrt{2}}.\label{supdiss}
\end{eqnarray}
For $\kappa<2f_c$, the superradiant phase exists for any regime of
coupling strength $\lambda \ge \lambda_c^-(\kappa)$, as shown in Fig.\ref{fig2}(a). In contrast, it only exists in the range $\lambda_c^-(\kappa)\le\lambda<\lambda_c^+(\kappa)$ for $\kappa>2f_c$, where
\begin{eqnarray}
\lambda_c^{\pm}(\kappa) = \sqrt{\frac{\kappa
f_c}{2}}\sqrt{1+\frac{4(\omega_0-\omega+
Jz_0)^2}{(2f_c\mp\kappa)^2}}.
\end{eqnarray}
We find that the line of commensurate excitation number $\mathcal{N}_c=1$ (shown in Fig.\ref{fig2}(a) as black dashed line) is given by,
\begin{eqnarray}
\frac{f_c\kappa}{2}+\frac{f_c(\omega-\omega_0-
Jz_0)^2}{2\kappa\left(\frac{3}{2}-\frac{f_c}{\kappa}\right)^2} =
\lambda^2, \label{commensurate_line_dissipation}
\end{eqnarray}
which meets the boundary between NP$_1$ and ${\rm SR}$ phase at
the critical point $\{\lambda_{f_c}^*,\kappa^*\}$,
\begin{eqnarray}
\lambda^*_{f_c} = \sqrt{f_c^2+\lambda^{*2}},\quad \kappa^* = 2f_c,
\end{eqnarray}
where $\lambda^{*}=(\omega-\omega_0-Jz_0)/2$ corresponds to the critical point in the non-dissipative limit $\kappa\rightarrow 0,\,f_c\rightarrow 0$, $f_c/\kappa \to \lambda^2/[2(\omega-\mu_0-Jz_0)^2]$, as discussed in Appendix.\ref{appa}.

We note that the transition between the superradiant and ${\rm NP}_1$ steady states corresponds to a dynamical transition, which at the classical level can be viewed as
a pitchfork bifurcation, where both $n^*$ and $s_z^*$ plays the role of the order parameter (see Fig.\ref{fig2}(b)). As a result of the $U(1)$ symmetry, the fixed points after such a bifurcation lie on a ring of radius $\sqrt{2n^*}$ on the complex plane of corresponding
photon field $\alpha^*$. The corresponding phase diagram exhibiting the ${\rm NP}$ and the SR phases is shown in Fig.\ref{fig2}(a).

\subsection{Excitation spectrum}
\label{excitation_spectrum}
To investigate the stability of the different phases, we perform a linear stability analysis of the EOM (Eq.\eqref{EOM_dissipation}) around the fixed points representing the steady states. We consider fluctuation at each site around the fixed points as,
\begin{eqnarray}
\mathbf{X}_i(t) &=&
\mathbf{X}^*+\delta\mathbf{X}_i(t)\\
{\bf X}_i^T &=& \{\psi_i, n_i, \phi_i, s_{iz}\}\nonumber\\
{\bf X}^{*T} &=& \{\psi^*, n^*, \phi^*, s_{z}^*\}\nonumber
\label{xdef}
\end{eqnarray}
and retain only the linear terms in fluctuation $\delta {\bf X}_i$. We substitute $\delta\mathbf{X}_i(t) = \int d^d r \exp(i\vec{q}.\vec{r_i})\,\delta\mathbf{X}_q(t)$ and obtain the set of linear equations for fluctuations in momentum space. These are
solved using the substitution
\begin{eqnarray}
\delta\mathbf{X}_q(t)=\exp(\mathcal{E}(q) t)\delta\mathbf{X}_q.
\label{xsol}
\end{eqnarray}
Thus, we construct the corresponding fluctuation matrix, whose eigenvalues $\mathcal{E}(q)$ yields the dispersion of the collective excitations. In the absence of dissipation, these excitations calculated from Linear stability analysis can alternatively be obtained by using the Holstein Primakoff approximation \cite{HP_transformation}, as discussed in Appendix.\ref{appa}.

In the dissipative system, $\mathcal{E}(q)$ can be complex and the stability of a steady state, in our notation, is ensured by the condition ${\rm Re}(\mathcal{E}(q))<0$. For the
unstable state, the degree of instability can be quantified by the exponent $\Lambda_{\rm I}={\rm max} ({\rm Re}\,\mathcal{E}(q))$ with ${\rm Re}(\mathcal{E}(q))>0 $. In contrast, ${\rm Im}(\mathcal{E}(q))$ yields the collective oscillation frequencies of the corresponding stable phases, similar to the excitations discussed for the non-dissipative case.

From the linear stability analysis, we find that the NP$_1$ phase becomes unstable when $\lambda\ge \lambda_c^-(\kappa)$ and undergoes a dissipative transition to a stable SR phase for which ${\rm Re}({\mathcal E}(q)) <0$ and we dub this phase as SR$_{\rm S}$.
We now investigate the nature of the excitations of the stable SR phase as obtained from ${\rm Im}(\mathcal{E}(q))$. Similar to the non-dissipative case, SR$_{\rm S}$ phase displays a gapless Goldstone mode along with a gapped Higgs mode, however, their dispersion relations exhibit intriguing behavior with strength of dissipation as discussed in Appendix.\ref{appb}.
Next, we focus on the excitations along the commensurate line joining the normal phase NP$_1$ at $\lambda_{f_c}^*$ (see black dashed line in Fig.\ref{fig2}(a)). At the critical point $\{\lambda_{f_c}^*,\kappa^*\}$, the excitation spectrum is given by,
\begin{eqnarray}
		\mathcal{E}^{\pm}(q) = \frac{1}{2}\left[-2f_c\pm i\omega_q\pm
		\sqrt{4f_c^2-\omega_q(4\lambda^*+\omega_q)}\right],
		\label{critical-point}
\end{eqnarray}
where we consider the cut $q_i=q$ in the Brillouin zone and $\omega_q = Jz_0(1-\cos q)$. In the low momentum ($q\rightarrow0$) regime, both the Higgs and Goldstone modes become degenerate with quadratic dispersion,
\begin{eqnarray}
		\mathcal{E}_{\rm Im}^{\pm}(q) \simeq \frac{Jz_0}{4}q^2,
\end{eqnarray}
as shown in Fig.\ref{fig:4}(c) of the Appendix.\ref{appb}. Therefore, in the presence of the incoherent decay and pumping processes, the dynamical critical exponent of the transition at the critical point $\{\lambda_{f_c}^*,\kappa^*\}$ is $z=2$, in contrast to the non-dissipative TC lattice, where the dispersion is linear with the critical exponent $z=1$ (see Eq.\eqref{comgold1} of the Appendix.\ref{appa}). Interestingly, the linear relativistic dispersion relation of the Higgs mode can be recovered even in the presence of dissipation at a special point on the commensurate line (see Appendix.\ref{appb} for details).

The SR$_{\rm S}$ phase becomes unstable at $\lambda=\lambda_{\rm I}$ as shown in Fig.\ref{fig2}(a); at this point, ${\rm Re}({\mathcal E}(q))$ changes sign. For $\lambda > \lambda_{\rm I}$, we denote the unstable SR phase by ${\rm SR_U}$. For $\kappa>2 f_c$, ${\rm SR_U}$ meets the unstable normal phase NP$_0$ (denoted by NP$_0^{\rm U}$ in Fig.\ref{fig2}(a)) at $\lambda_c^+(\kappa)$ shown by the red dashed-dotted line in Fig.\ref{fig2}(a). The normal phase NP$_0$ becomes stable above a critical coupling strength $\tilde{\lambda}_c(\kappa)$ given by,
\begin{eqnarray}
\tilde{\lambda}_c(\kappa) = \sqrt{\frac{\kappa f_c}{2}}\sqrt{1+\frac{4(\omega_0-\omega- Jz_0)^2}{(2f_c-\kappa)^2}},
\end{eqnarray}
which is shown in the phase diagram given in Fig.\ref{fig2}(a) and this stable normal phase is dubbed as NP$_0^{\rm S}$. The variation of $n^*$ and $s_z^*$ corresponding to the SR phase as a function of $\lambda$ for a fixed value of $\kappa,f_c$ is depicted in Fig.\ref{fig2}(b). Note that, even though the total excitation number is not conserved in the presence of dissipation, it is possible to obtain a line of fixed excitation number in the stable regime of the SR phase, since its steady state value does not change in time.

The presence of the unstable ${\rm SR_U}$ and ${\rm NP_0^U}$ states in an extended parameter regime (white region of the phase diagram in Fig.\ref{fig2}(a)) indicates the absence of stable homogeneous steady state solutions in this regime. This leads to natural questions regarding the presence of other possible inhomogeneous steady states and also the fate of several variables, both microscopic and collective (site-averaged), in those states. To address these questions, next, we study the quench dynamics of the model.

\section{Dissipative quench dynamics}
\label{dissdyn}

In this section, we explore the non-equilibrium quench dynamics of the system in a parameter regime for which the homogeneous steady state corresponds to unstable ${\rm SR}$ or ${\rm NP}_0$. The aim of this study will be to delineate the nature of these steady states. We start from an initial configuration of variables $n_i$, $s_{iz}$, $\psi_i$ and $\phi_i$, which describe the initial state of the system and study the time evolution of the density matrix by numerically solving the EOM given in Eq.\eqref{EOM_dissipation}; this is equivalent to studying quench dynamics of the system. We study such dynamics for a finite number of sites arranged in the form of one dimensional lattice with periodic boundary
condition. Thus we solve the classical equation of motion without assumptions of a homogeneous solution {\it i.e.}, by allowing the possibility of site-dependent variation of dynamical variables $n_i(t)$, $s_{iz}(t)$, $\psi_i(t)$ and $\phi_i(t)$. As we shall see,
this will be crucial for understanding the nature of the unstable steady states and the long time dynamical behavior of this system. In what follows, we shall study the behavior of the decorrelator in Sec.\ref{deco} and discuss the details of the steady state reached
due to this dynamics in Sec.\ref{steady}. In Sec.\ref{scar1}, we discuss the violation of ergodicity in this system due to the emergence of the scarring phenomenon. For all numerical simulations, we consider a 1D lattice with a total number of sites $N_0=1000$. 

\subsection{Probing instability from Decorrelator}
\label{deco}

In the coupling regime of the unstable steady state, it is natural to expect that fluctuations will dominate the dynamics of the microscopic variables. Due to these fluctuations, the driven system is expected to  exhibit chaotic dynamics in spite of the presence of dissipation. To probe such instability and chaotic behavior, in recent years the classical version of the out-of-time correlator, namely the decorrelator has been widely used \cite{Decorr1,Decorr2,Decorr3,Decorr4,Decorr5,Decorr6}. It is defined as
\begin{eqnarray}
D(i,t) &=& 1-\langle\langle {\bf s}_i^a\cdot {\bf
s}_i^b\rangle\rangle,
\end{eqnarray}
where $a$ and $b$ denote two copies of the initial configuration of the spin variables. The configuration $b$ is chosen to be slightly perturbed from $a$ at the central site $i=i_0$; all other sites $i \ne i_0$ have identical initial configurations for all variables. Here
$\langle\langle...\rangle\rangle$ indicates the average over different initial configurations chosen uniformly from the available phase space. It is expected that for a completely chaotic system, the perturbation will display a light cone like spreading leading to the growth of the decorrelator. In contrast, for a stable dissipative steady state, the dynamical variables at each site quickly converge to the corresponding steady state values, restricting the spread of the decorrelator, which remains close to its initial value. This allows the dynamics of $D(i,t)$ to distinguish between unstable and stable regimes.
\begin{figure}[t]
	\includegraphics[width=\linewidth]{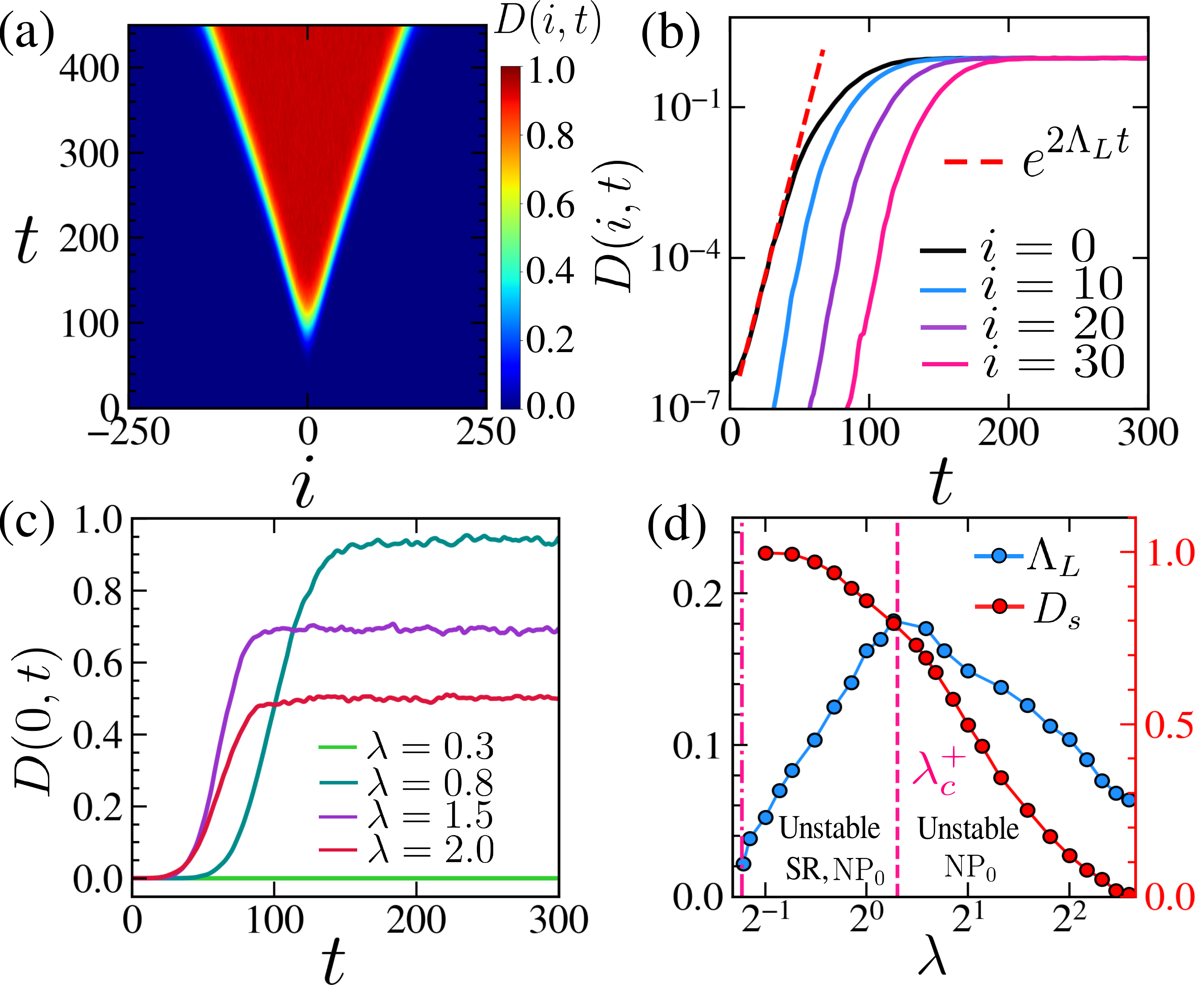}
	\caption{{\it Chaotic behavior from the dynamics of decorrelator in 1D TC lattice.}
		(a) Light cone like spatiotemporal spreading of decorrelator $D(i,t)$.
		(b) Exponential growth of $D(i,t)$ in time for different sites
		$i$ and the extraction of Lyapunov exponent $\Lambda_L$ from
		$D(i_0,t)$. (c) Long time saturation value $D_s$ of decorrelator
		$D(i_0,t)$ for different values of coupling strengths $\lambda$.
		(d) Variation of Lyapunov exponent $\Lambda_L$ and saturation value
		$D_s$ as a function of $\lambda$. The left axis corresponds to
		Lyapunov exponent $\Lambda_L$ and right axis indicates the values of
		$D_s$. The $\Lambda_L$ is maximum at the critical point
		$\lambda_{c}^+$ above which unstable SR does not exist.
		The figures (a,b) correspond to $\lambda=0.8$. In this and rest of the figures, we choose the total number of sites to be $N_0=1000$ and $10^4$ different random initial conditions to obtain any ensemble averaged quantity.
		Parameter chosen: $\kappa=0.3$.}
	\label{fig3}
\end{figure}

The numerical results obtained for $D(i,t)$ are shown in Fig.\ref{fig3}. In the unstable regime (see Fig.\ref{fig2}(a)), the spatiotemporal growth of the decorrelator $D(i,t)$ displays a light cone like spreading of the initial perturbation. This is in contrast
to its behavior in the stable regime, where $D(i,t)$ remains close to zero for almost all $i$. Such a stark contrast in the growth of the perturbation in terms of the decorrelator can be used to identify the unstable regime. This feature becomes apparent in Fig.\ref{fig3}(c) where $D(i_0,t)$ shows qualitatively different behavior for $\lambda=0.3$, corresponding to the stable superradiant phase SR$_{\rm S}$.
\begin{figure*}
	\includegraphics[width=0.8\linewidth]{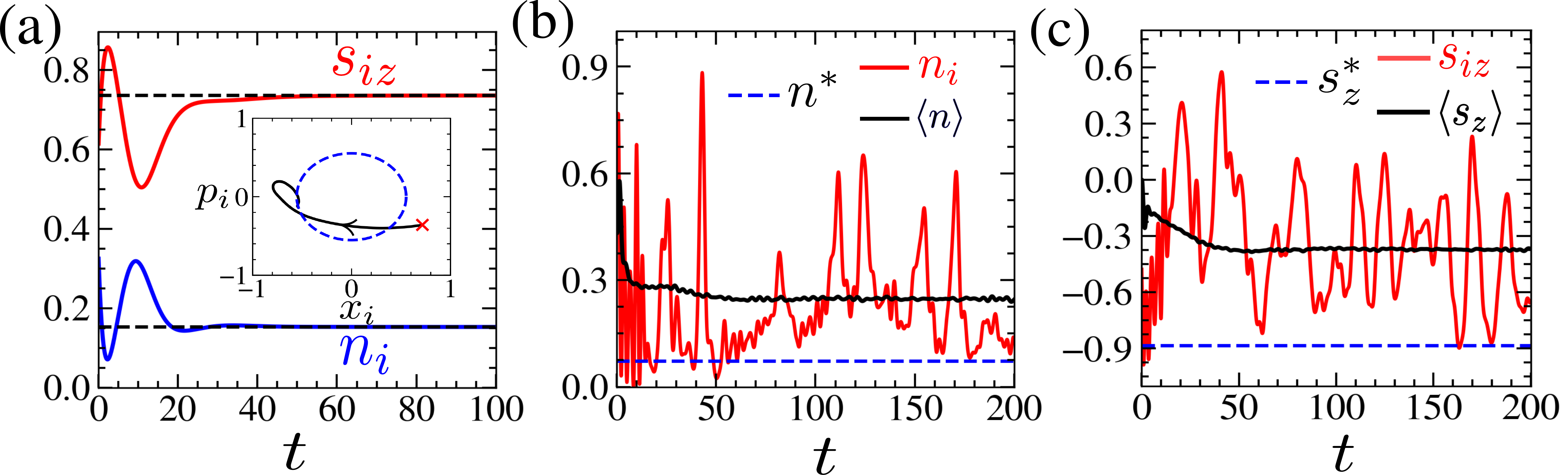}
	\caption{{\it The dynamics of $n_i$ and $s_{iz}$ at any arbitary site $i$ in
		the stable and unstable regimes of the SR phase in the 1D lattice.} (a) The relaxation dynamics of the local photon number $n_i$
		and atomic population $s_{iz}$ for the stable regime of SR phase corresponding to $\lambda = 0.3$. The dashed black lines are corresponding steady state values. The inset of (a) shows the relaxation dynamics towards the ring of fixed points in the $x-p$ plane. In the unstable regime, the dynamics of (b) $n_i$
		and (c) $s_{iz}$ shown by red color for $\lambda=1.0$. The dynamics of
		$\langle n\rangle(t)$ and $\langle s_z\rangle(t)$ averaged over all
		sites, shown by black lines in (b,c), exhibit much smaller
		fluctuation indicating the quasi-steady state. The photon number
		$n^*$ and atomic population $s_{z}^*$ of the steady state SR$_{\rm U}$ are marked by
		dashed blue line, which are significantly deviated from the
		corresponding values of the quasi steady state. See text for
		details. We consider the decay rate $\kappa=0.3$ for all figures.}
	\label{fig4}
\end{figure*}

In the unstable regime, we find that $D(i,t)$ starts growing from zero after a certain time $t_0$ and saturates to a particular value $D_s$ at a long time. Due to the light cone spreading of the decorrelator, $t_0$ increases linearly as $\sim|i-i_0|$; the specific value of $t_0$ depends on the butterfly velocity. The instability of the driven system can be quantified from the exponential growth of the decorrelator $D(i,t)\sim \exp(2\Lambda_L t)$, where $\Lambda_L$ is the Lyapunov exponent which can be extracted appropriately from
the decorrelator of the central site $i=i_0$, as shown in Fig.\ref{fig3}(b). The variation of $\Lambda_L$ with the coupling strength $\lambda$ is depicted in Fig.\ref{fig3}(d). It is evident from Fig.\ref{fig3}(d) that $\Lambda_L$ attains a maxima at $\lambda= \lambda_{c}^+$, {\it i.e.}, at the boundary of the unstable SR and ${\rm NP}_0$ phases, shown by the dashed line in Fig.\ref{fig3}(d). Also, $\Lambda_L \simeq 0$ for $\lambda<\lambda_{\rm I}$ (shown by the dashed-dotted line in Fig.\ref{fig3}(d)) indicates the absence of chaos in the stable SR and NP$_1$ phases.

The long time saturation value of the decorrelator, $D_s$, is another indicator of the chaotic behavior and its value for several representative coupling strengths $\lambda$ and for $\kappa=0.3$ is shown in Fig.\ref{fig3}(c). The variation of $D_s$ with $\lambda$
is depicted in Fig.\ref{fig3}(d). We find that $D_s$ monotonically decreases from unity as one increases $\lambda$ and vanishes at the point where the stable NP$_0$ phase appears. We note that $D_s$ decreases much more rapidly with $\lambda$ in the unstable ${\rm
NP}_0$ phase compared to that in the unstable SR phase.

The study of $D(i,t)$ clearly points out the qualitative distinction between the behavior of dynamics and the nature of steady states for the stable and unstable regimes. In the next
section, we study these features of steady states in the unstable regime using collective site-averaged variables, which provides a deeper insight into the nature of these steady states.

\subsection{Emergence of a quasi-steady state}
\label{steady}

In the last section, we found chaotic behavior in the regime where the homogeneous ansatz for the microscopic variables yields an unstable steady state; such chaotic behavior is manifested through the exponential growth of the decorrelator. It is expected that in this
regime, the dynamical variables at each site will experience strong fluctuations. However, the site-averaged collective variables, for which these fluctuations may be small, may provide further information about the nature of the steady state reached at long
times in the course of the dynamics. We therefore study the dynamics of the site averaged collective quantities,
\begin{eqnarray}
\langle n\rangle(t) = \frac{1}{N_0} \sum_{i=1}^{N_0} n_i(t), \quad
\langle s_z\rangle(t)= \frac{1}{N_0}\sum_{i=1}^{N_0}
s_{iz}(t),\,\,\,
\label{avnsz}
\end{eqnarray}
where $N_0$ is the total number of sites in the cavity. In addition, we shall also study the time-averaged values $\langle n\rangle(t)$ and $\langle s_z\rangle(t)$ given by,
\begin{eqnarray}
\overline{ \langle n\rangle} = \frac{1}{T} \int_{t_0}^{T+t_0}\, dt\,
\langle n\rangle(t),  \overline{\langle s_{z}\rangle} =
\frac{1}{T} \int_{t_0}^{T+t_0}\, dt\, \langle s_z\rangle(t)  \quad
\label{tavnsz}
\end{eqnarray}
where $t_0$ is chosen to be larger than the typical transient time.

For the long time dynamics, the deviation of the on-site physical quantities $\{n_i(t), s_{iz}(t)\}$ from the corresponding values of fixed point $\{n^{*}, s_{z}^*\}$ is a measure of fluctuation around the steady state and serves as a marker to distinguish between
stable and unstable steady states. For the parameter regime which corresponds to stable SR (${\rm SR}_S$) phase, both these dynamical variables at any particular site ($n_i$ and $s_{iz}$) as well as their site-averaged values ($\langle n\rangle (t)$ and $\langle
s_z\rangle(t)$) approach to the corresponding fixed points within a time scale $\tau_0 \sim 1/{\rm Max}[{\rm Re}(|\mathcal{E}(q)|)]$. Such dynamics of $n_i(t)$ and $s_{iz}(t)$ is
shown for an arbitrary site $i$ in Fig.\ref{fig4}(a). The dynamics of photons in the $x-p$ plane, shown in the inset of Fig.\ref{fig4}(a) for the same parameter regime, quickly approaches a steady state on the ring of fixed points. These plots indicate the expected gradual approach of $n_i(t)$ and $s_{iz}(t)$ to their steady state values; the other on-site variables such as $\psi_i(t)$, $s_{ix}(t)$ and $s_{iy}(t)$ also behave in an identical manner.

In contrast, in the unstable regime (${\rm SR_U}$ phase), the dynamics do not follow any steady state in the strict sense, as depicted in Fig.\ref{fig4}(b) and (c); this is particularly obvious when one monitors the dynamics of local variables $n_i(t)$ and
$s_{iz}(t)$ (red curves), exhibiting large fluctuations. However, the long-time dynamics of the collective site-averaged variables $\langle n\rangle(t)$ and $\langle s_z\rangle(t)$ attain certain quasi-steady value with small fluctuation around it (depicted in Fig.\ref{fig4}(b,c)). This quasi-steady state reached in the long time dynamics is distinctly different from the unstable ${\rm SR_U}$ phase, which is evident from Fig.\ref{fig4}(b,c).

\begin{figure}[t]
\includegraphics[width=\linewidth]{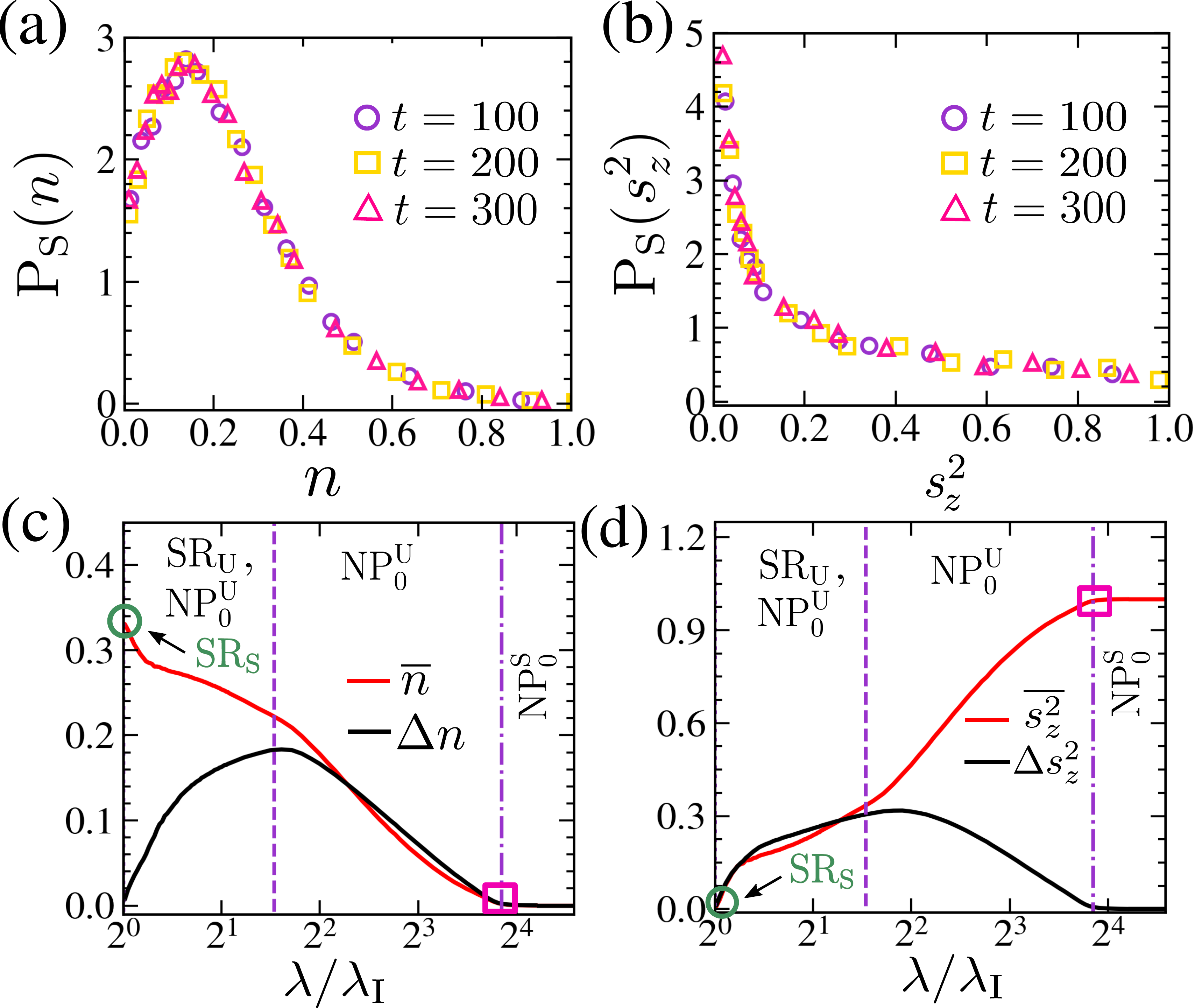}
\caption{{\it Statistical features of the emergent quasi steady state:} The spatial distribution of the microscopic variables (a)
$n_i$ and (b) $s_{iz}^{2}$ at $\lambda = 1.0$, displaying a stationary form at long
time. The variation of time average value of the mean (red solid)
and standard deviation (black solid) of the quasi steady state with
the coupling strength $\lambda$ (base 2 log scale) is shown for (c)
$n_i$ and (d) $s_{iz}^2$. Here, the violet color dotted, dashed,
dash-dotted lines correspond to $\lambda = \lambda_{\rm I}, \lambda_c^+$ and
$\tilde{\lambda}_c$ respectively. Stable (unstable) superradiant and the normal phases are denoted by SR$_{\rm S}$ (SR$_{\rm U}$) and NP$^{\rm S}_0$ (NP$^{\rm U}_0$), respectively.
The quasi steady state interpolate between stable SR$_{\rm S}$ and NP$_0^{\rm S}$ phases. For all plots, $\kappa=0.3$.}
\label{fig5}
\end{figure}

Although there are possibilities of other attractors in the unstable domain, we numerically checked that for arbitrary initial configurations the spatially averaged collective variables always approach the same quasi-steady state values. Thus the behavior of
$\overline {\langle n\rangle}$ and $\overline {\langle s_z\rangle}$ is similar to that due to a thermalization mechanism, where the memory of initial configuration is lost and the macroscopic thermodynamics quantities attain a steady value, even though the microscopic variables rapidly fluctuate. We emphasize that this phenomenon occurs in the presence of dissipation and is in complete contrast to that of the usual stable steady state expected to be reached in long time dynamics of the dissipative systems.

\begin{figure}[b]
	\includegraphics[width=0.7\linewidth]{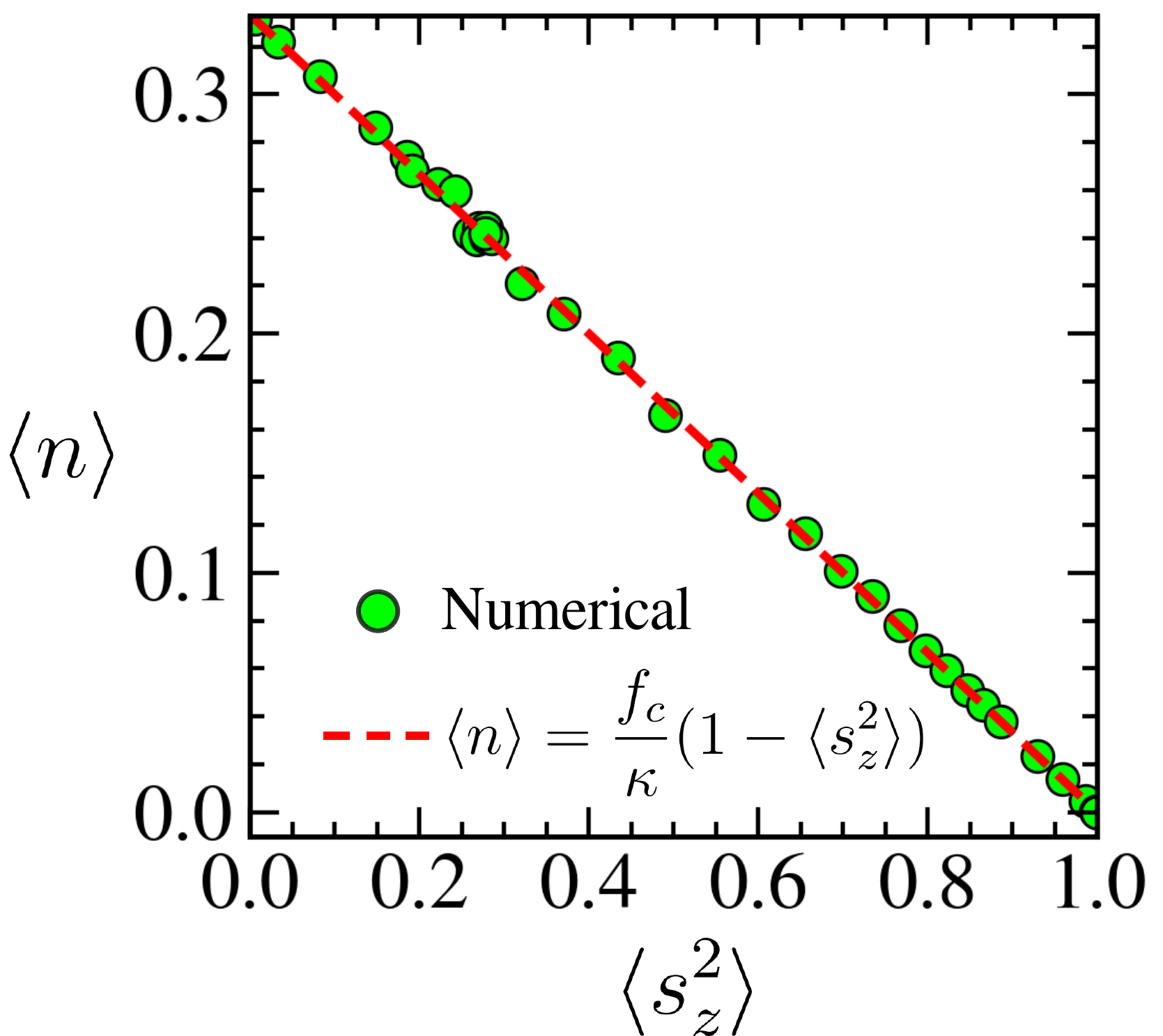}
	\caption{The linear relation between the collective variables
		$\langle n\rangle, \langle s_z^2\rangle$ corresponding to the quasi
		steady state. We choose the decay rate $\kappa=0.3$.}
	\label{fig6}
\end{figure}

Since the collective variables $\langle n\rangle, \langle s_z\rangle$ attain a steady value, we investigate the site distribution of the dynamical variables $n_i$ and $s_{iz}$, which, interestingly, attain a stationary distribution after a sufficiently long time (See Fig.\ref{fig5}(a,b)), although the individual microscopic variables fluctuate in time. Even though the distribution of photon resembles the shape of the Poisson distribution, as shown in Fig.\ref{fig5}(a), the main characteristic of Poissonian statistics $(\Delta n)^2\sim
\langle n\rangle$ is violated for certain regime of $\lambda$, reflecting sub- or super-Poissonian distributions. In fact, our analysis shows that the spatial distribution of the photon number in this regime can be fitted to a distribution function $(a+x^b) \exp[-c x^d]$, where $a$, $b$, $c$ and $d$ are constants. From these distributions of $n_i$, $s_{iz}$, we obtain the mean values $\langle n\rangle,\langle s_z^2\rangle$ and corresponding
standard deviations $\Delta n,\Delta s_z^2$, which can be used to characterize the quasi-steady states for these collective variables, as illustrated in Fig.\ref{fig5}(c,d). By the manipulation of the EOMs given in Eq.\eqref{EOM_dissipation}(a,b), we find that the site averaged quantities satisfy the relation
\begin{eqnarray}
\langle n\rangle = \frac{f_c}{\kappa}(1-\langle s_z^2\rangle),
\end{eqnarray}
which is obeyed by the collective variables obtained from the corresponding distributions, as shown in Fig.\ref{fig6}. Remarkably, the mean values $\langle n\rangle$ and $\langle
s_z\rangle$ of this new emergent steady state interpolate between the stable SR$_S$ and NP$_0$ phases, which is shown in Fig.\ref{fig5}(c,d). It is also evident from Fig.\ref{fig5}(c,d) that the spatial fluctuations of both these variables attain a maximum value near the coupling $\lambda_c^+$, where the Lyapunov exponent also attains its maximum value (see Fig.\ref{fig3}(d)). This behavior reflects the underlying chaos, as discussed in the previous section.

\begin{figure*}[t]
	\includegraphics[width=\linewidth]{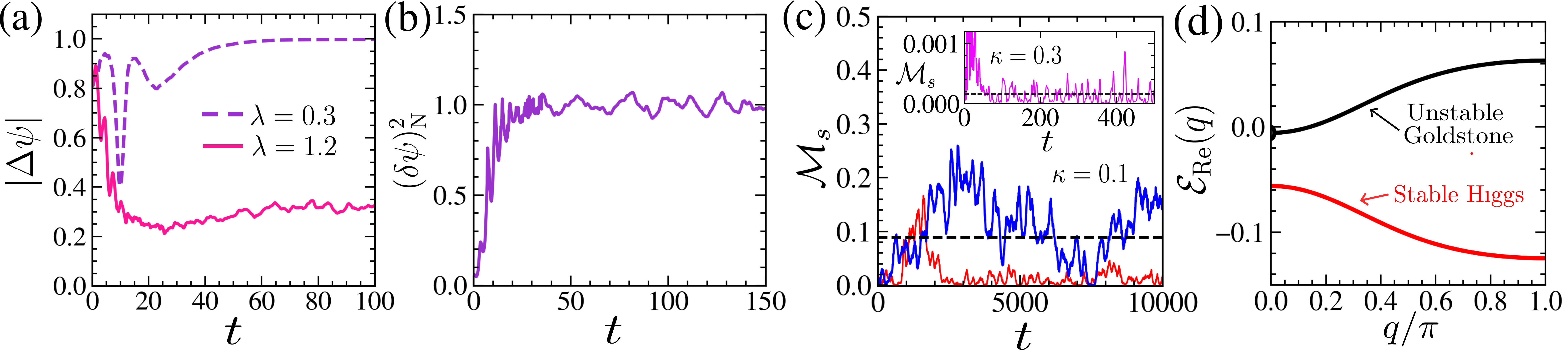}
	\caption{{\it The non-equilibrium dynamics in the unstable regime of
			superradiant phase {\rm SR}$_{\rm U}$}. (a)
		Dynamics of the phase fluctuation between neighbouring sites, described by $|\Delta\psi|$ in
		the stable regime of SR phase at $\lambda=0.3$ and for unstable
		regime $\lambda = 1.2$ with $\kappa=0.3$. (b) Normalized photon phase fluctuation
		$(\delta\psi)^2_{\rm N}$ for $\lambda = 1.2$. (c) Time evolution of the spin
		order parameter $\mathcal{M}_s$ at $\kappa=0.1$ and $\lambda=2.0$, starting from two different
		initial conditions. The black dashed line marks the ensemble average value of the order parameter $\langle\langle \mathcal{M}_s\rangle\rangle$. The inset shows the dynamics of $\mathcal{M}_s$ at $\kappa=0.3$ with $\lambda=2.0$. (d) The real part of the excitation spectrum $\mathcal{E}_{\rm Re}(q)$ of SR$_{\rm U}$ phase, exhibiting unstable Goldstone mode.
	}
	\label{fig7}
\end{figure*}
To gain more information about such quasi-steady states, we investigate the phase fluctuation of photons between the neighboring sites, which can be captured from the following quantity,
\begin{equation}
\Delta \psi (t) = \frac{1}{N_{0}z_0}\sum_{i,\delta}
[\cos(\psi_{i}(t) -\psi_{i+\delta}(t))], \label{phase_fluc}
\end{equation}
where, $\delta$ denotes nearest neighbor of $i^{\rm th}$ site and $z_0$ is the coordination number. $\Delta \psi$  has also been used to probe quantum mechanical fluctuation of the relative phase between two sites of the Bose Josephson junction \cite{Oberthalar}. In the stable superradiant state SR$_{\rm S}$, the phase $\psi_i$ takes the same value at all sites as a consequence of U(1) symmetry breaking and consequently the relative phase difference $|\Delta\psi(t)|$ rapidly attains the steady value unity, depicted in Fig.\ref{fig7}(a). On the other hand in the unstable regime, $|\Delta \psi(t)|$ becomes much smaller than
unity, exhibiting temporal fluctuation (solid line in the Fig.\ref{fig7}(a)).

The phase $\psi_i$ of photon not only has a temporal fluctuation but also fluctuates at different sites. The spatial fluctuation of phase over different sites can be computed from the quantity
\begin{eqnarray}
    (\delta \psi)^2_{\rm N}(t) = (\delta \psi)^2(t)/(\delta \psi)^2_{{\rm max}},
\end{eqnarray}
where $(\delta \psi)^2(t)= \sum_i [\psi_i(t) -\langle\psi\rangle(t)]^2/N_0$ is the phase variance, $\langle \psi\rangle(t)$ is the mean value of the phase at time $t$, and $(\delta \psi)^2_{{\rm max}}=\pi^2/3$ is the maximum value of the fluctuation corresponding to a
uniform distribution of phases. As shown in Fig.\ref{fig7}(b), $(\delta \psi)^2_{\rm N}(t)$ exhibits rapid growth and approaches to unity, indicating large spatial fluctuations at long times.

Similar to the photon field, the breaking of U(1) symmetry for the spin degree of freedom can be captured from the collective variable
\begin{eqnarray}
 \mathcal{M}_s = \langle s_+\rangle\langle s_-\rangle,
\end{eqnarray}
which acts as the order parameter and becomes finite when the system is in the broken U(1) symmetry phase. As evident from Fig.\ref{fig7}(c), $\mathcal{M}_s$ remains small without approaching a steady value in the unstable SR phase, unlike the collective quantities $\langle n\rangle$ or $\langle s_z\rangle$. Moreover, the time average value of $\mathcal{M}_s$ starting from each initial condition is different from the corresponding ensemble average value $\langle\langle \mathcal{M}_s\rangle\rangle$, indicating a non-ergodic behavior, as depicted in Fig.\ref{fig7}(c). Moreover, $\mathcal{M}_s$ decreases as we approach towards the stable NP$_0$ phase with increasing the parameters $\kappa$ or $\lambda$ (see inset of Fig.\ref{fig7}(c)).
\begin{figure}[t]
	\includegraphics[width=\columnwidth]{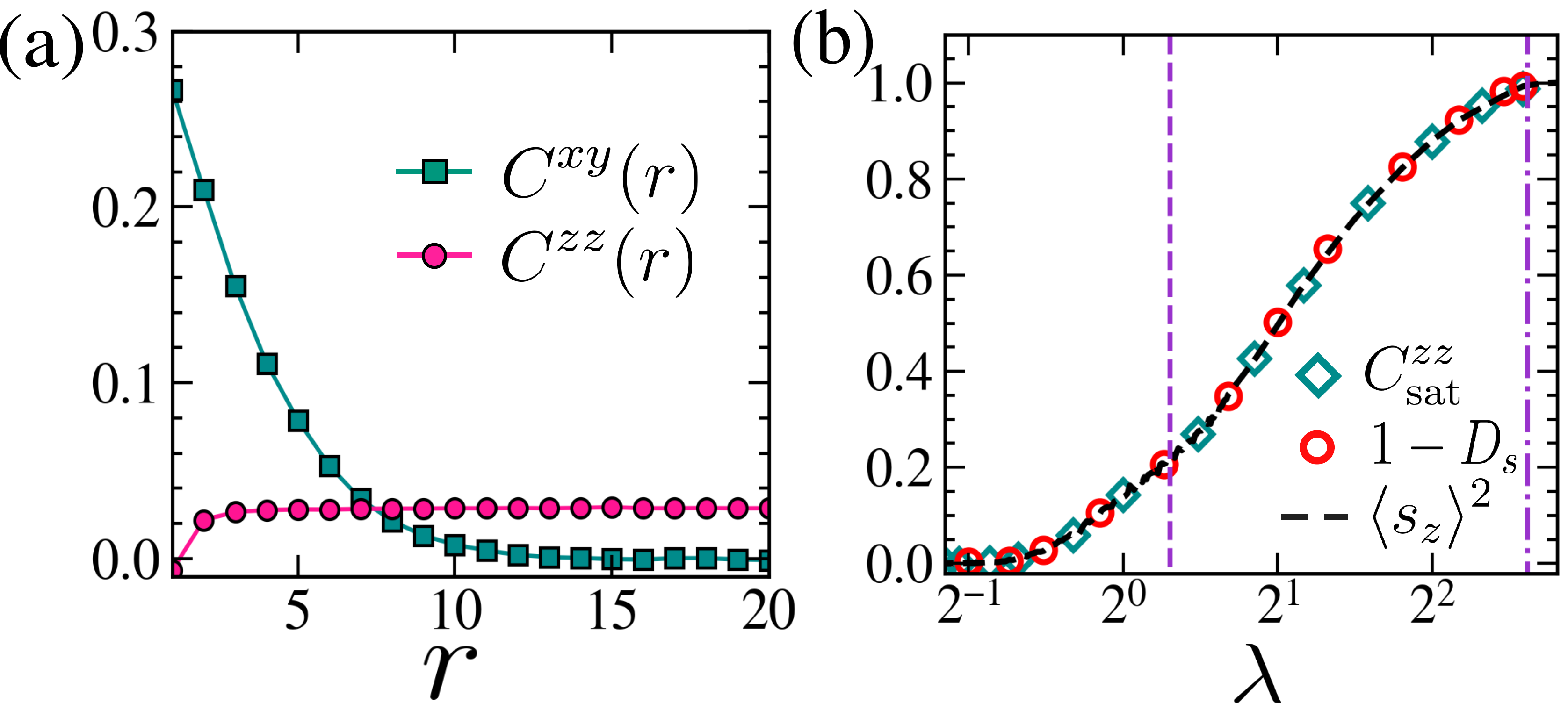}
	\caption{{\it Anisotropy in the behavior of the correlation functions of the spin degree of freedom.} (a) Correlation functions in $x$-$y$ plane $C^{xy}(r)$ and in the $z$ direction $C^{zz}(r)$, as a function of distance $r$ for coupling strength $\lambda=0.7$.
	(b) Comparison of the saturation value of the correlation function $C^{zz}_{\rm sat}$ in the $z$ direction, average value of the $z$ component of the spin $\langle
	s_z\rangle^2$ and saturation of the decorrelator $D_s$ as a function of
	coupling strength $\lambda$. The violet dashed (dashed-dotted) line represents the
	coupling strength $\lambda_c^+$ ($\tilde{\lambda}_c$). Parameter
	chosen: $\kappa=0.3$. }
	\label{fig8}
\end{figure}

We note that in the SR$_{\rm U}$ phase the Goldstone mode becomes dynamically unstable (see Fig.\ref{fig7}(d)), which possibly triggers such phase fluctuation of the photons (as shown by $(\delta \psi)^2_N$ and $|\Delta \psi|$) as well as the fluctuation of the order parameter $\mathcal{M}_s$ in the $x-y$ plane.

Next, we analyze the dynamics of two-spin correlation functions. To investigate the ordering in $x-y$ and $z$ direction, we compute
\begin{subequations}
\begin{eqnarray}
&&C_{ij}^{xy}  = \langle\langle s_{ix}s_{jx}+s_{iy}s_{jy}\rangle\rangle\\
&&C_{ij}^{zz}  = \langle\langle s_{iz}s_{jz}\rangle\rangle
    \end{eqnarray}
\end{subequations}
where $\langle\langle..\rangle\rangle$ denotes averaging over an ensemble of random initial conditions. The behavior of the correlation functions with distance $r=|i-j|$ are shown in Fig.\ref{fig8}(a), which reveals an anisotropic behavior in $z$ and $x$-$y$ direction. Both these correlation functions can be fitted with the functional form
\begin{eqnarray}
C^{a b}(r) &\sim&  C_{\rm sat}^{a b}+c\exp(-r/\xi^{a b}),
\end{eqnarray}
where $(a,b)$ takes $(z,z)$ or $(x,y)$ for the out-of-plane or in-plane spin correlations, respectively. Here, $C_{\rm sat}^{ab}$ and $\xi^{ab}$ represent the saturation value and the correlation lengths respectively. 
Our numerical evaluation of these correlators finds a clear anisotropy, where $C_{ij}^{zz}$ quickly attains its saturation value $C_{\rm sat}^{zz}\simeq \langle s_z\rangle^2$, describing the ordering in the $z$ direction while $C_{ij}^{xy}$ decays much more slowly and saturates to a small value. This anisotropic nature of spin correlation in the $x-y$ plane and $z$ direction is also reflected in the behavior of the correlation length $\xi^{xy}\gg\xi^ {zz}$. Such spin correlation can also be identified from the saturation value of the decorrelator $D_s\approx
1-C_{\rm sat}^{zz}$, since $C_{\rm sat}^{xy}\approx 0$, as shown in Fig.\ref{fig8}(b). This analysis also elucidates the physical significance of the saturation value of the decorrelator, where $1-D_s$ quantifies the ordering in the $z$ direction of the spin system.
\begin{figure*}
	\includegraphics[width=\textwidth]{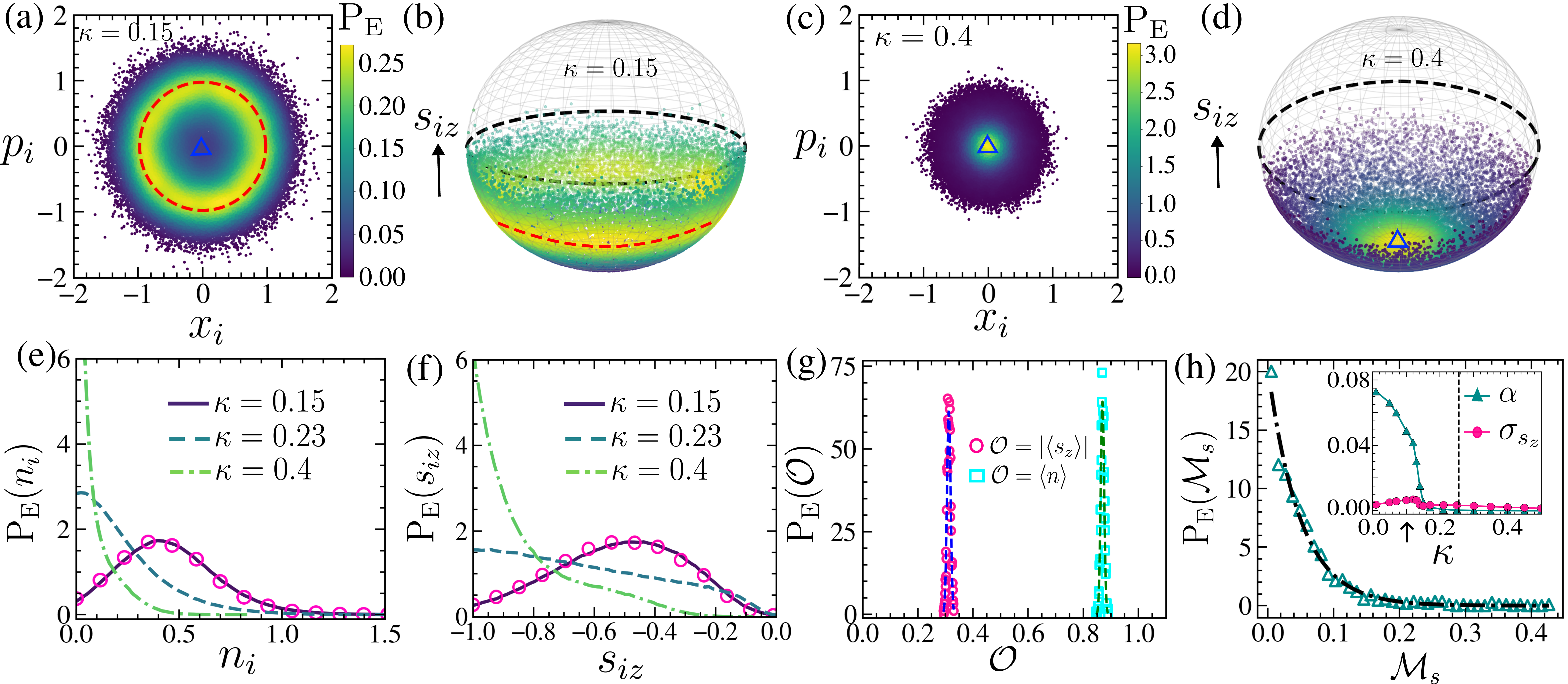}
	\caption{{\it Effect of underlying attractors on the overall dynamical behavior}. The phase space distribution of (a,c) photon field  ${\rm P_E}(x_i,p_i)$ in $x_i$-$p_i$ plane and (b,d) spins on the Bloch sphere ${\rm P_E}(\theta_i,\phi_i)$ at $i$th site over an ensemble of random initial conditions. The color scale in these figures indicates the probability density of the phase space points. The black dashed line on the Bloch sphere indicates the equator.  The red dashed line in (a,b) and blue triangle in (c,d) represent the steady state of the unstable SR and NP$_0$ phases. The decay rate corresponding to (a,b) is $\kappa=0.15$ and (c,d) is $\kappa=0.4$. The distributions P$_{\rm E}$ of the microscopic variables (e) $n_i$ and (f) $s_{iz}$ over an ensemble of random initial states for different decay rates $\kappa$, which exhibit a well defined most probable value. The pink circles in (e, f) represent the respective distributions over lattice sites P$_{\rm S}$ for $\kappa=0.15$. The distribution of the collective variables (g) $\langle n\rangle,\langle s_z\rangle$ (fitted by the Gaussian function) and (h) order parameter $\mathcal{M}_s$ (fitted by an exponential function) over the different initial conditions for $\kappa=0.1$. The inset of (h) shows the variation of the fitting parameter $\alpha$ and the width $\sigma_{s_z}$ of the
	distribution of $\langle s_z\rangle$ with $\kappa$. We choose the coupling strength $\lambda=2.0$ for all the figures. }
	\label{fig9}
\end{figure*}

Overall our analysis in this section points out a couple of unusual properties of the dynamics in the parameter regime which corresponds to the unstable steady state (${\rm SR_U}$). First, the long-time dynamics in this regime leads to a steady state, where microscopic variables such as $n_i$, and $s_{iz}$ experience strong fluctuations while the collective site-average variables $n$ and $s_z$ attain a definite steady state value irrespective of the initial condition. This is a reminiscence of the ergodic behavior, which is usually not expected in a semiclassical system with dissipative dynamics. Second, we identify the existence of two classes of collective variables in the system with qualitatively different behavior. The first class of these variables include $n$ and $s_z$; these variables and their correlation functions exhibit small fluctuation around a well-defined steady state value which is reached within a relatively short time. In the parlance of dynamical systems, these collective variables approach a fixed point of the phase space during time evolution. The second class includes $\mathcal{M}_s$ and $\Delta \psi$; these variables experience strong fluctuations at all times and do not have a unique steady state value even at long times. These collective variables retain the memory of the initial condition and flow to a fixed surface in the phase space. This behavior is also to be contrasted with the usual dissipative dynamics of a semiclassical driven dissipative system. Although here we have presented the results for one dimensional lattice, we have verified that the similar steady state also exists in two dimensions.

\subsection{Ergodicity and collective scarring phenomenon}
\label{scar1}

In this section, we study the properties of the emergent collective steady state. To this end, we time evolve the dynamical variables following the Eq.\eqref{EOM_dissipation}, starting from an ensemble of random initial conditions and study the final phase space
configuration after a sufficiently long time. We note that each site of the array of cavities is statistically equivalent; consequently, we may focus on the final phase space distributions P$_{\rm E}(n_i)$ and P$_{\rm E}(s_{iz})$ of the dynamical variables $\{n_i,s_{iz}\}$ at any arbitrary site $i$. Analyzing these distributions at a specific site yields physical insights into the nature of the attractor and the way dynamical variables are distributed on it. In this context, note that for the array of cavities with many degrees of freedom, the final phase-space configuration of all variables is expected to be enormously complicated; consequently, the statistical equivalence of sites is crucial for a tractable analysis of the steady state.

For an arbitrary site, we first study the distribution of the photon field in $x_i$-$p_i$ plane and spins over the Bloch sphere, as shown in Fig.\ref{fig9}(a-d). In the parameter regime, which corresponds to the unstable superradiant phase ${\rm SR_U}$, we find that the distribution of the photon field in $x_i$-$p_i$ plane takes the shape of a ring and is sharply peaked around it, as shown in Fig.\ref{fig9}(a). This is similar to the characteristics of the ${\rm SR_U}$ state although the state reached in dynamics is different from it. Moreover in spin space, the distribution of ${\vec s}_i$ is localized on a circle in the lower hemisphere of the Bloch sphere with a most probable value of $s_{iz}$ corresponding to the steady state of the unstable SR phase, as illustrated in Fig.\ref{fig9}(b). As one approaches the stable NP$_0$ phase by varying $\kappa$ for a fixed $\lambda$, the nature of the distribution of ${\vec s}_{i}$ changes above a certain $\kappa$ so that the spins are concentrated near the south pole of the Bloch sphere ($s_{z}=-1$) and the photon fields are distributed around the origin ($x=p=0$), as depicted in Fig.\ref{fig9}(d,c) respectively. Notably, the distributions P$_{\rm E}(n_i)$ and P$_{\rm E}(s_{iz})$ of these dynamical variables are concentrated around the fixed point of the unstable NP$_0$ phase. This change in the behavior of distributions of the microscopic variables is shown for several representative values of $\kappa$ in Fig.\ref{fig9}(e,f).

Similarly, we can study the distribution of the site-averaged collective variables
or order parameters, describing the properties of the photon field and the spin system corresponding to the quasi-steady state. The distributions of $\langle n\rangle,\,\langle s_z\rangle$ exhibit a very sharply peaked distribution, similar to a delta function, centered at the values that correspond to the steady state values of the respective quantities, shown in Fig.\ref{fig9}(g). From a dynamical perspective, this suggests the presence of a
fixed point for these quantities within the multi-dimensional phase space. Such a distribution corresponds to the ergodic behavior of these collective variables since it reflects the complete loss of memory of the initial configuration in the steady state. In contrast, the distribution of the order parameter $\mathcal{M}_s$, shown in Fig.\ref{fig9}(h), follows an exponential distribution P$_{\rm E}(\mathcal{M}_s) = \exp(-\mathcal{M}_s/\alpha)/\alpha$. Here $\alpha$ denotes the mean (and the variance) of the distribution; a larger $\alpha$ indicates a wider distribution. Interestingly, within a certain regime of the phase diagram for small $\kappa$, the width of the distribution of $\mathcal{M}_s$ is significantly large. This behavior clearly indicates a fixed surface within the phase space, signifying that the memory of the initial configuration is still retained in the steady state. However, the most probable value of this order parameter remains vanishingly small.

\begin{figure}[b]
\includegraphics[width=0.6\linewidth]{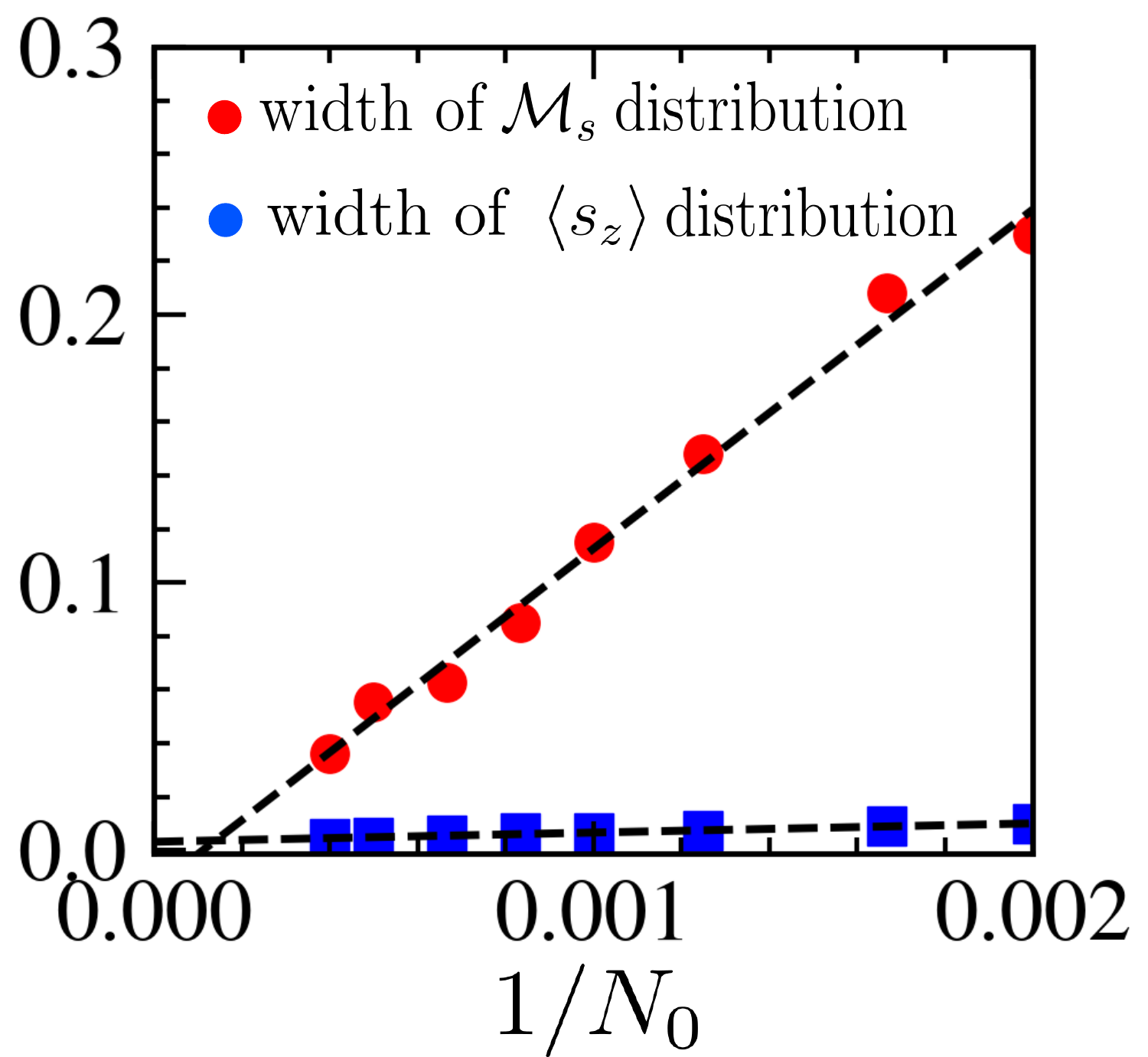}
\caption{Plot for the width of the distribution of $\mathcal{M}_s$ and $\langle s_z\rangle$ as a function of $1/N_0$ for $\kappa=0.1$ and $\lambda=2.0$, showing the distinction between the two classes of collective variables at finite $N_0$. 
}
\label{fig10}
\end{figure}

We note that the width of the distribution $\alpha$, shows a weak finite-size effect, with its value decreasing as the number of lattice sites $N_0$ increases, which is shown in Fig.\ref{fig10}. We note that although the $1/N_0$ extrapolation seems to indicate that
all variables retain ergodicity in the thermodynamic limit, for any experimentally relevant
size of this array (for example, $N_0 \le 2000$), there is a distinction in the behavior of these two classes of variables. Thus the collective variables related to the phase degrees of freedom, such as $\mathcal{M}_s$, can lead to weak breaking of ergodicity at finite $N_0$.

From the distribution of microscopic variables, specifically ${\rm P_E}[n_i],{\rm P_E}[s_{iz}]$ discussed previously, we have verified that the ensemble
average of $n_i$ and $s_{iz}$ at any arbitrary site correctly yields
$\sum_{n_i} {\rm P_E}[n_i] n_i= \langle n\rangle$ and $\sum_{s_{iz}} {\rm P_E}[s_{iz}]
s_{iz}= \langle s_z \rangle$, where $\langle n\rangle$ and $\langle s_z \rangle$ represent the site averaged values of respective variables at the steady state. This confirms ergodicity in terms of {\it the equivalence between ensemble averaging and time averaging}.
Furthermore, we have checked that the ensemble distributions of these variables at any arbitrary site match with the distributions ${\rm P_S}[n_i]$ and ${\rm P_S}[s_{i z}]$ at the steady state for a single realization, such that ${\rm P_E}[n_i]={\rm P_S}[n_i]$ and ${\rm P_E}[s_{i z}] = {\rm P_S}[s_{i z}]$ (see Fig.\ref{fig9}(e,f)). As a consequence, the ensemble average of any collective (thermodynamic) quantity associated with the photon number and atomic excitations will be identical to the time-averaged value of the corresponding quantity. This provides a definitive confirmation of ergodicity.

\begin{figure}[t]
	\includegraphics[width=\columnwidth]{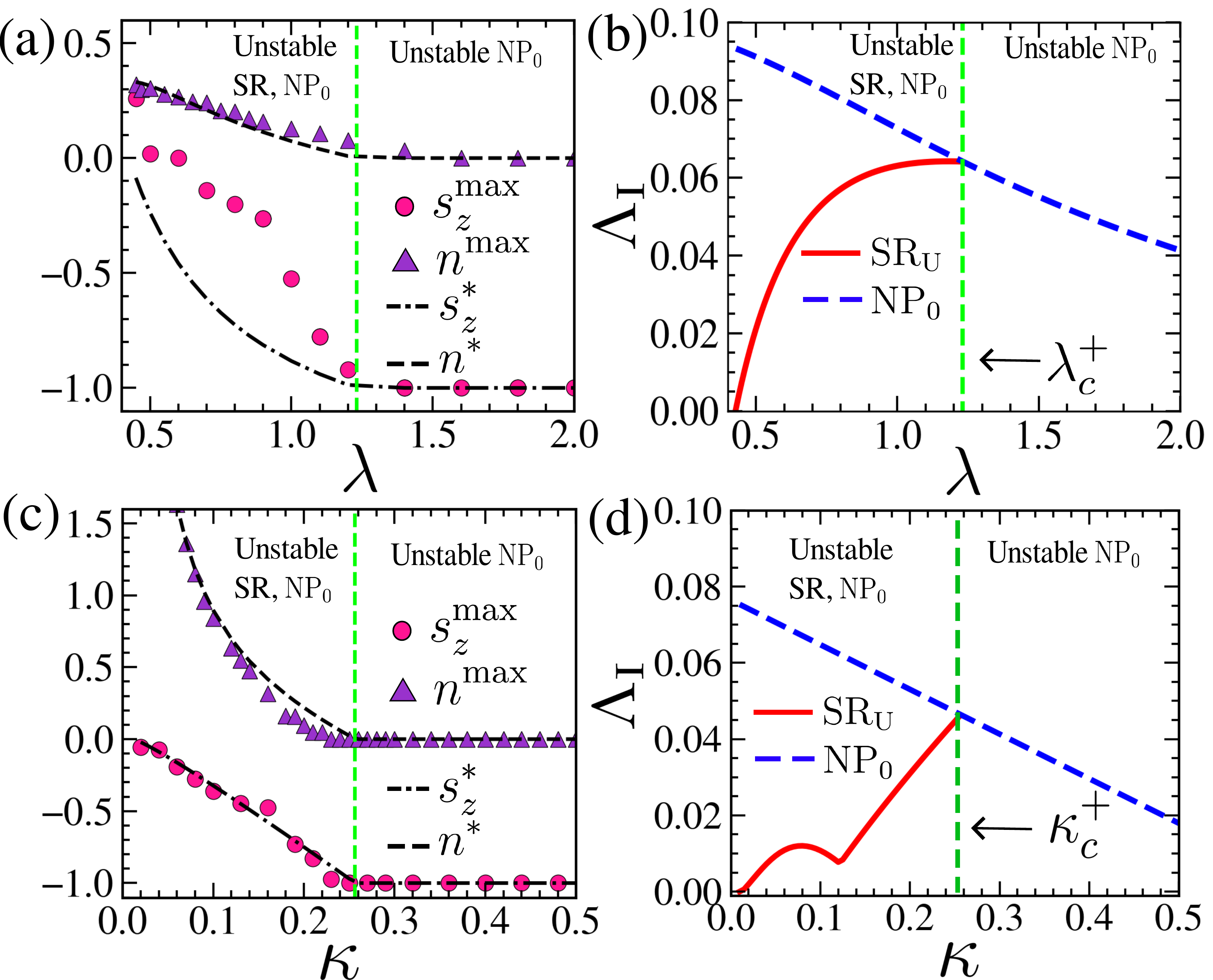}
	\caption{{\it The collective scarring phenomenon and the
		reminiscence of unstable steady states.} The variation of the most
		probable value $n^{\rm max}, s_z^{\rm max}$ obtained from the
		distribution of microscopic variables $n_i,s_{iz}$ as a function of
		(a) $\lambda$ for $\kappa=0.3$ and (c) $\kappa$ for $\lambda=2.0$,
		which are close to the corresponding unstable steady state values
		marked by black dashed and dashed dotted lines. The green dashed
		line in panel  a(c) indicates the coupling $\lambda_c^+$ ($\kappa_c^+$),
		above which unstable SR phase do not exist. The instability exponent
		$\Lambda_{\rm I}$ for the unstable SR phase (solid red line) and
		NP$_0$ phase (dashed blue line) as a function of (b) $\lambda$ with
		$\kappa=0.3$ and (d) $\kappa$ for $\lambda=2.0$. }
	\label{fig11}
\end{figure}

A closer analysis of the ensemble distributions of microscopic variables $\{n_i,s_{iz}\}$, unveils yet another intriguing feature of this quasi-steady state. We observe that the distributions of $\{n_i,s_{iz}\}$ are peaked around $n^{\ast}$ or $s_z^{\ast}$, {\it i.e.}, the values corresponding to the unstable steady states, which can be either SR or ${\rm NP}_0$, depending on the coupling strength. As a result, the most probable value of the photon number $n^{\rm max}$ and atomic excitation $s_z^{\rm max}$ follow the unstable steady states (as depicted in Fig.\ref{fig11}(a,c)), manifesting the influence of these unstable fixed points on the statistical nature of the quasi steady state. Away from the stable NP$_0$ phase, the most probable values of these variables are attracted towards the unstable SR phase. As we approach the stable NP$_0$ phase, a crossover takes place above a certain coupling strength. Consequently, both $n^{\rm max},s_z^{\rm max}$ are drawn towards the NP$_0$ phase (see Fig.\ref{fig11}(a,c)), even though it remains unstable.

The aforementioned crossover of the most probable values of the
dynamical variables can also be understood from the degree of instability
of the unstable steady states, which can be quantified from their
instability exponent $\Lambda_{\rm I}={\rm max}({\rm Re}\,\mathcal{E}(q))$. In the regime of small $\kappa$ or $\lambda$, where both the unstable SR and NP$_0$ coexist, the
dynamical variables are attracted towards the SR phase, which has a smaller
instability exponent. On the contrary, as we approach the stable NP$_0$
phase, its instability exponent decreases and it attracts the
dynamical variables, as explained in Fig.\ref{fig11}(b,d). We have
also verified that the site distributions of the microscopic variables
and their most probable values also exhibit similar behavior, as shown in Fig.\ref{fig9}(e,f). This analysis reveals the dramatic influence of the unstable steady states on the distribution of the microscopic variables even when there is underlying chaos, as evident from the light cone dynamics of the decorrelator. This example
elucidates how the microscopic variables are attracted towards the unstable steady states and their crossover even in the presence of chaos.

We call this phenomenon as {\it collective scarring}, owing to its similarity with the single particle scarred states observed in chaotic quantum systems; in other words, the
reminiscence of the unstable steady state can be identified from the accumulation of the phase space density corresponding to the scarred state. The impact of unstable dynamics on scarring phenomena has been studied in single particle \cite{E_Heller}, as well as in collective quantum systems \cite{Bose_Hubbard_scar,BJJ_Sudip,D_mondal_scar,KCT}.
Notably, this collective scarring phenomenon is exhibited by the dynamical variables, which obey ergodicity, in contrast, quantum scars lead to weak violation of ergodicity. In a typical closed chaotic system, scarred trajectories often are not statistically significant
(for exceptions to this see, this Ref.\cite{uscar1}); however, in the present context of systems with dissipation, it seems that all initial conditions for $n_i$ and $s_{iz}$ lead to such trajectories.

Finally, we address the issue of the quantum steady state of this open atom-photon array in the dynamically unstable regime, which is described by the density matrix. For large magnitude of spin $S\gg1$, semiclassically, the density matrix can be described by the Glauber-Sudarshan P representation \cite{P_rep_1,P_rep_2},
\begin{eqnarray}
	\hat{\rho} = \int{\rm P}(\alpha_i,\mathcal{Z}_i) \ket{\alpha_i,\mathcal{Z}_i}\bra{\alpha_i,\mathcal{Z}_i} d^2\alpha_id^2\mathcal{Z}_i
	\label{P_rep}
\end{eqnarray}
where ${\rm P}(\alpha_i,\mathcal{Z}_i)$ is the probability distribution of the classical variables $\alpha_i$, $\mathcal{Z}_i=\tan(\theta_i/2)\exp(\iota\phi_i)$, corresponding to the photon and the spin degree of freedom of each cavity. 
This classical distribution ${\rm P}(\alpha_i,\mathcal{Z}_i)$ can be obtained within the truncated Wigner approximation (TWA), involving an ensemble of classical trajectories \cite{TWA_1,TWA_2,A_M_Rey_TWA}. 
The initial conditions of the trajectories are sampled from a Gaussian distribution around a classical phase space point to simulate quantum fluctuations.
Semiclassically, for large but finite $S$, a phase space point can be represented by a product coherent state  $\ket{\Psi_c} = \prod_i\ket{\alpha_i}\otimes \ket{\mathcal{Z}_i}$ for the photon and spin degree of freedom. The classical initial points are sampled from the corresponding Husimi distribution. 
%
%
\begin{figure}
	\includegraphics[width=\columnwidth]{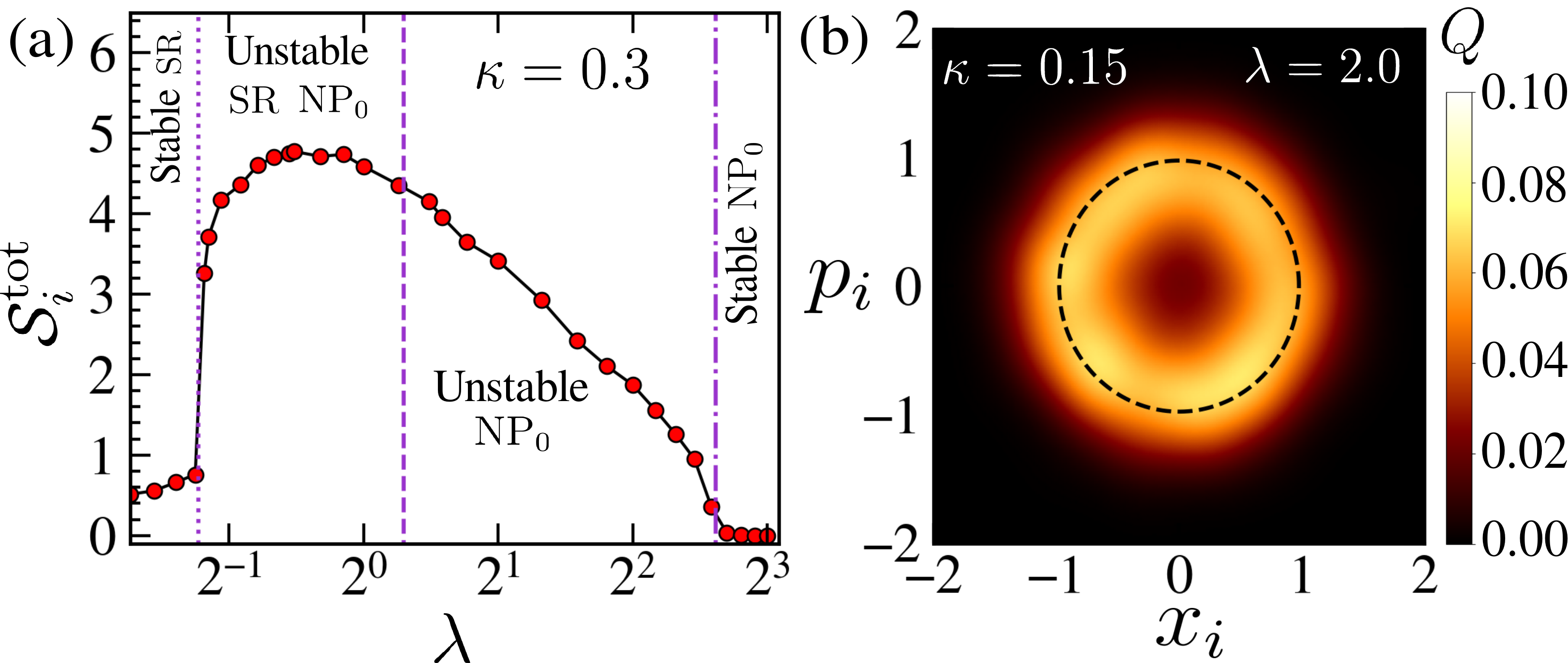}
	\caption{ {\it Properties of the semiclassical density matrix} : (a) Variation of the von Neumann entropy $\mathcal{S}^{\rm tot}_i$ corresponding to the steady state density matrix $\hat{\rho}_i^{\rm ss}$ at an arbitary site  with coupling strength $\lambda$ for $\kappa=0.3$. (b) Husimi distribution $Q$ of the photon field on $x-p$ plane in the regime of unstable SR phase for $\kappa=0.15,\lambda=2.0$. Black dashed circle represents the FPs of unstable SR phase. We consider $S=20$.}
	\label{fig12}
\end{figure}
Neglecting the stochastic fluctuations, which are suppressed as $1/\sqrt{S}$ for large $S$, the time evolution of the classical variables is obtained by the EOM, given in Eq.\eqref{EOM_dissipation}. Within TWA, the physical quantities like photon number and spin polarization can be obtained by averaging over the ensemble of trajectories.
Notably, even in this method, the dynamical variables follow the same stationary distributions P$_{\rm E}(n_i)$, P$_{\rm E}(s_{zi})$ after a sufficiently long time, irrespective of the initial choice of phase space point.
From the stationary distribution of phase space variables at each cavity,
corresponding steady-state density matrix $\hat{\rho}^{\rm ss}_i$ can be constructed using Eq.\eqref{P_rep}, which in turn yields the total density matrix 
in the product form $\hat{\rho} = \prod_i \hat{\rho}_{i}$ within the mean-field approximation. From $\hat{\rho}^{\rm ss}_i$, we compute the von Neumann entropy per site $\mathcal{S}^{\rm tot}_i={\rm Tr}(\hat{\rho}^{\rm ss}_i\ln\hat{\rho}^{\rm ss}_i)$ and its variation with coupling strengh.
%
%
As evident from Fig.\ref{fig12}(a), the emergent steady state in the chaotic regime has significantly higher entropy compared to the states corresponding to stable fixed points (FPs), such as the SR and NP phases, clearly distinguishing the nature of the steady state in the respective dynamical regimes.
Moreover, the variation of entropy with coupling strength resembles the behavior of the saturation value of the decorrelator, which quantifies the degree of chaos.
To better understand the photon field in the steady state of the dynamically unstable regime, we compute the reduced density matrix $\hat{\rho}^{\rm ss}_{\rm Ph}$ by tracing out the spin degree of freedom.
Corresponding Husimi distribution $Q(\alpha,\alpha^*)=\frac{1}{\pi}\bra{\alpha}\hat{\rho}^{\rm ss}_{\rm Ph}\ket{\alpha}$ of the photon, projected in the $x-p$ plane exhibits a smeared phase space density localized around the circle of classical FPs of unstable SR phase (see Fig.\ref{fig12}(b)), illustrating the collective scarring effect.
%


\section{Discussion}
\label{conc}
In this work, we have studied a coupled atom-photon system described by the Tavis-Cummings Hubbard model in the presence of dissipation and pumping. Our main objective is to understand the nature of the steady states realized in these systems. To this end, we have studied the systems in the limit of large $N$ and performed a semiclassical EOM analysis to obtain our results.

Our analysis reveals the presence of unstable homogeneous steady states in the system for a wide range of parameter values. This results from the instability of the low-lying excitations around such steady states (for example, Goldstone modes for the SR phase); in this regime, one finds the existence of an incoherent atom-photon mixture with phase fluctuation. Such states are qualitatively different from typical expected stable dissipative steady states. Our computation of the decorrelator shows ballistic light cone spreading of excitations in the unstable regime; this clearly distinguishes it from its stable counterparts. These two classes of states can also be distinguished by the steady state value of the decorrelator; for the unstable steady state, it reaches a finite value while for the stable ones, it remains close to zero.

To understand the nature of the driven systems better, we study quench dynamics in this unstable regime and unravel the existence of a quasi-steady state. In this state, the microscopic variables show wide temporal fluctuations; in contrast, the site-averaged collective variables reach a steady state value. These variables can be divided into two distinct classes. The first class of variables such as $n$ and $s_z$ display ergodic behavior and do not retain any memory of the initial condition. In the parlance of a dynamical system, this corresponds to a fixed point of the quench dynamics; their distribution mimics a delta function. In contrast, the second class of variable, such as $\mathcal{M}_s$, retains the memory of the initial condition and shows an exponential distribution; this corresponds to a fixed surface in the phase space.

We find that the width $\alpha$ of this exponential distribution for the second class of variables is a slowly varying function of the system size $N_0$; it approaches zero in the thermodynamic limit. In this limit, it seems that all collective variables display ergodic behavior. Interestingly, the spatio-temporal fluctuations of the microscopic
variables persist in this limit; this constitutes an emergence of an ergodic steady state in a dissipative environment in the thermodynamic limit.
We also point out that for any experimentally realizable array size of such cavities (say, $N_0 \le 2000$), the distinction between these two classes of variables is clear. Thus this phenomenon corresponds to weak ergodicity breaking in a finite-sized atom-photon array and constitutes an example of a quasi-steady state where two classes of variables show distinct behavior. To the best of our knowledge, this phenomenon has not been reported earlier in the context of driven dissipative systems.

Finally, we find that in the unstable regime, the variables of the ergodic class are attracted towards the unstable homogenous steady states. Consequently, the steady state distributions of these variables are peaked around these values signifying the influence of the unstable homogeneous steady state on the statistical properties of these quasi ergodic states. This feature is reminiscent of the scarring at the classical level, where phase space density accumulates around the unstable dynamical state, which we refer to as {\it collective scarring}. However, we distinguish this from the quantum scars in usual many-body systems, which occupy a vanishingly small fraction of the many-body Hilbert space and are associated with ergodicity violation; on the contrary, here the collective scarring phenomenon does not violet ergodicity as the collective variable {\it i.e.} average photon number and atomic excitation correspond to their steady state values. 
 

To understand the nature of the quantum steady state in the chaotic regime, we also perform truncated Wigner approximation, which also reveals the formation of an incoherent fluid of photons in this regime. Interestingly, the entropy of this emergent steady state is significantly higher compared to the stable superradiant and normal phases, which is the distinguishing feature of this state in the chaotic regime.

The experimental platforms that can verify our work require the realization
of standard coupled cavities. Such systems, for a few cavities coupled by photons,
have been realized using different experimental platforms such as superconducting qubits \cite{maierref}, NV spin centers \cite{nvcref}, cavity QED setup \cite{baksic1,cavref1} and coupled quantum dots \cite{qdots}. A detailed account of such systems can be found in \cite{jcrefs0}. Although the scalability of such arrays is an experimental challenge, it is conceivable that such arrays may be realized in the future. We predict that in such a driven dissipative array, there will be a wide parameter regime where upon quenching the atom and photon number on any given site will show significant temporal fluctuations while their site average values will display a time-independent steady state value. Moreover, the behavior of these variables will be different from the site-average photon phase or superfluid density; the latter will have a finite range of steady state values depending on the initial condition.

To conclude, we have studied the non-equilibrium phases and dynamics of a dissipative atom-photon system described by the TCH model in the semiclassical limit.
The present work unravels the emergence of a unique quasi-steady state where the collective variables attain a steady value while the local microscopic variables exhibit spatio-temporal fluctuations. We identified one class of dynamical variables that exhibits both ergodicity and collective scarring, retaining the signature of unstable steady states, while another class only weakly violates ergodicity in finite-size systems.
Our work not only demonstrates the existence of an experimentally achievable incoherent photonic fluid but also sheds light on the ergodicity in dissipative systems.

\section{Acknowledgements}
KS thanks A. Sen for discussion and SERB for support through project
JCB/2021/000030. SS and DM thank Iacopo Carusotto and Sudip Sinha for discussions.

\appendix

\section{Semiclassical analysis of the model without dissipation}
\label{appa}

In this appendix, we study the steady state phases of the Hamiltonian $\hat{{\mathcal H}}$ (Eq.\eqref{Hamiltonian} of the main text) and their excitation spectra in the absence of dissipation. As discussed in the main text, in the limit of the large magnitude of the spins $S\gg 1$, the system can be described classically. The corresponding classical Hamiltonian $\mathcal{H}_c$ in the grand canonical ensemble and the excitation number $\mathcal{N}_c$ are given in Eq.\eqref{hameq} and \eqref{excitationeq} respectively. The corresponding EOM can be obtained from the Hamilton's equations as follows,
\begin{eqnarray}
	\dot{n}_i &=& -\lambda\sqrt{2n_i}\sqrt{1-s^2_{iz}}\sin(\phi_i-\psi_i)\nonumber\\
	&-&2J\sum_{\delta}\sqrt{n_in_{i+\delta}}\sin(\psi_i-\psi_{i+\delta})\nonumber\\
	\dot{\psi}_i &=& (\omega-\mu)+\frac{\lambda\sqrt{1-s_{iz}^2}}{\sqrt{2n_i}}\cos(\phi_i-\psi_i)\nonumber\\
	&-&J\sum_{\delta}\sqrt{\frac{n_i}{n_{i+\delta}}}\cos(\psi_i-\psi_{i+\delta})\nonumber\\
	\dot{\phi}_i &=& (\omega_0-\mu)-\frac{\lambda\sqrt{2n_i}\,s_{iz}}{\sqrt{1-s_{iz}^2}}\cos(\phi_i-\psi_i)\nonumber\\
	\dot{s}_{iz} &=&
	\lambda\sqrt{2n_i}\sqrt{1-s_{iz}^2}\,\sin(\phi_i-\psi_i),
	\label{EOM}
\end{eqnarray}
where the sum over $\delta$ represents sum over the nearest neighbors of the $i^{\rm th}$ site.  Next, we analyze these EOM to obtain the phases and excitation spectra of the system.

\subsection{Steady state and phase transition}
\label{Steady_state}

\begin{figure}[b]
	\centering
	\includegraphics[width=\columnwidth]{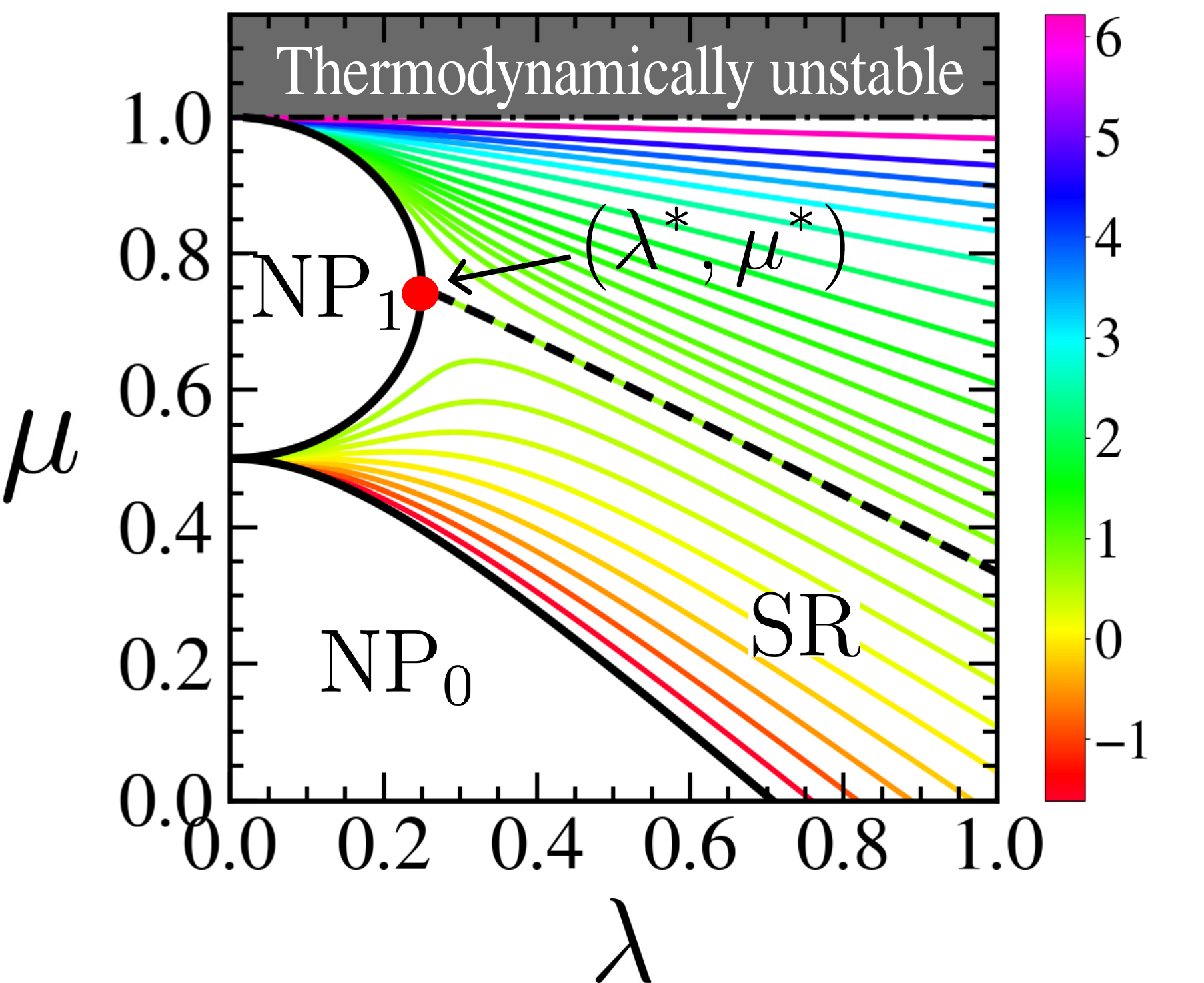}
	\caption{The phase diagram of the Tavis-Cummings lattice model for large
		$S$ and in the homogeneous limit. The fixed excitation number lines
		are plotted as color scale of $\ln\mathcal{N}_{\rm c}$. The line of
		commensurate excitation number ($\mathcal{N}_{\rm c}=1$), starting
		from the tip of the NP$_1$ lobe ($\lambda^*,\mu^*$), is marked by
		black dashed line. For this plot and in rest of the figures, we have
		chosen $\omega = 2.0$, $\omega_0=0.5$, and $Jz_0=1$.}
	\label{fig:1}
\end{figure}

In this subsection, we focus on the homogeneous solutions ($\psi_i=\psi,n_i=n,\phi_i=\phi,s_{iz}=s_z$) of the EOM given in Eq.\eqref{EOM}, where all the spins are aligned and the cavities are equally populated on each site. This leads to the dynamics of an effective single cavity TC model whose EOM are given by,
	\begin{eqnarray}
		\dot{n} &=&
		-\lambda\sqrt{2n}\sqrt{1-s^2_{z}}\sin(\phi-\psi)\nonumber\\
		\dot{\psi} &=&
		(\omega-\mu-Jz_0)+\frac{\lambda\sqrt{1-s_{z}^2}}{\sqrt{2n}}\cos(\phi-\psi)\nonumber\\
		\dot{\phi} &=&
		(\omega_0-\mu)-\frac{\lambda\sqrt{2n}\,s_{z}}{\sqrt{1-s_{z}^2}}\cos(\phi-\psi)\nonumber\\
		\dot{s}_{z} &=& \lambda\sqrt{2n}\sqrt{1-s_{z}^2}\,\sin(\phi-\psi)
		\label{single_cavity_EOM}
	\end{eqnarray}
where $z_0 = 2d$ is the coordination number of the lattice. It is
evident from Eq.\eqref{single_cavity_EOM} that the equations of motion depend only on the relative phase $\phi-\psi$ and not on individual phases. For such a homogeneous case, the classical Hamiltonian and the excitation number scaled by the number of lattice sites can be written as,
\begin{eqnarray}
		\mathcal{H}_{\rm c} &=& (\omega-\mu-Jz_0)n+(\omega_0-\mu)\,s_z\nonumber\\
		&+&\lambda\sqrt{2n}\sqrt{1-s_z^2}\cos(\phi-\psi)
		\label{classical_energy}\\
		\mathcal{N}_{\rm c}&=& \frac{n+1+s_z}{2}. \label{Conserved-number}
\end{eqnarray}
Note that, for $\mu>\omega-Jz_0$, there is no global minima in the system, since energy is minimized for $n\rightarrow\infty$. This can be easily seen from the classical energy given in Eq.\eqref{classical_energy}. Therefore, we restrict our discussion to the thermodynamically stable regime $\mu<\omega-Jz_0$. The thermodynamically unstable regime is marked in the phase diagram, given in Fig.\ref{fig:1}. We investigate the steady state phases of the system by indentifying the different fixed points ($\psi^{\ast},n^{\ast},\phi^{\ast},s_z^{\ast}$) of the EOM (Eq.\eqref{single_cavity_EOM}) and analyzing their stability for the total lattice, from the linear stability analysis, as discussed in the main text.

For a fixed conserved excitation number $\mathcal{N}_{\rm c}$, there exist two possible steady states corresponding to $\phi-\psi=0$ and $\pi$. These solutions determine different values of $\mu$. The state associated with $\phi-\psi=0$ lies in the thermodynamically unstable regime $\mu>\omega-Jz_0$, which we neglect in this work. Consequently, we only focus on $\phi-\psi=\pi$ for the steady state analysis.
In the case of $\phi-\psi=\pi$, we find that the spontaneous breaking of $U(1)$
symmetry leads to two distinct classes of continuous phase transitions
depending on the chemical potential $\mu$. For $\mu<\omega_0$ and
below the critical coupling strength $\lambda < \lambda_{c}$, where
\begin{eqnarray}
	\lambda_c &=& \sqrt{|(\omega_0-\mu)(\omega-\mu-Jz_0)|},
	\label{critlam}
\end{eqnarray}
the ground state represents the symmetry unbroken normal phase ${\rm NP}_0$ characterized by $n^{\ast}=0$ and $s_z^{\ast}=-1$, corresponding to
$\mathcal{N}_{\rm c}=0$ and energy $E =-(\omega_0-\mu)$. On the other hand, for $\lambda> \lambda_{c}$, a continuous set of stable superradiant (SR) phases emerges as the ground state, characterized by the nonzero photon number $n=n^{\ast}$ and $s_z=s_z^{\ast}$, where,
\begin{eqnarray}
	n^{\ast} &=&-\frac{(\omega_0-\mu)^2}{2\lambda^2}
	\left[\left(\frac{\lambda}{\lambda_c}\right)^4-1\right],\quad
	s_z^{\ast} = -\frac{\lambda_c^2}{\lambda^2}.\qquad
	\label{Superradiant}
\end{eqnarray}
Note that $n^{\ast}=0$ and $s_z^{\ast}=-1$ at $\lambda=\lambda_c$ indicates a continuous transition between these phases.
In the regime $\omega_0<\mu<\omega-Jz_0$, the normal phase NP$_1$ ($n^{\ast} =0$ and $s_z^{\ast}=1$) with excitation number $\mathcal{N}_{\rm c}=1$ and energy $E =\omega_0-\mu$ undergoes a transition at $\lambda=\lambda_c$ giving rise to a continuous ring of SR phases (Eq.\eqref{Superradiant}). The phase boundary in the $\mu-\lambda$ plane between the ${\rm NP}_1$ and the SR phases has two branches; both of these branches merge at the tip of the corresponding lobe (see Fig.\ref{fig:1}) given by,
\begin{eqnarray}
	\lambda^*=\frac{\omega-\omega_0-Jz_0}{2},\quad \mu^* =
	\frac{\omega+\omega_0-Jz_0}{2}. \label{tip-lobe}
\end{eqnarray}
The lines of conserved excitation number in $\mu-\lambda$ plane can
be achieved by substituting the steady state value of the superradiant
phase in Eq.\eqref{Conserved-number}. The line of commensurate
excitation number ($\mathcal{N}_{\rm c} = 1$) is given by,
\begin{align}
	\mu_c(\lambda) =
	\frac{1}{3}\left[3\mu^*+2\lambda^*-\sqrt{\lambda^{*2}+3\lambda^2}\right],
	\label{Commensurate_line}
\end{align}
which meets the phase boundary between the ${\rm NP}_1$ and the SR
phases at $\{\lambda^*,\mu^*\}$. The phase diagram in $\mu-\lambda$
plane, depicting the above phases and transitions between them is
shown in Fig.\ref{fig:1}. The contours in Fig.\ref{fig:1}
represent curves of fixed ${\mathcal N}_c$. The phase transition
across the tip of the lobe between the ${\rm NP}_1$ and the SR
phases exhibits different critical behavior than other points on the
phase boundary; this point shall be addressed below. We note that
the phase diagram is similar to that of the Bose-Hubbard model, exhibiting superfluid-insulator
transition \cite{th1,th2,th3,th4}.


\subsection{Excitation spectrum} \label{exspec}

For $S\gg1$, the excitations of the Tavis-Cummings lattice model can
be analyzed semiclassically by considering $1/S$ to be a small
parameter.  We apply a linear shift to the cavity mode as
$\tilde{a}_i=a_i+\beta_i\sqrt{S}$, so that it describes the quantum
fluctuations around the classical photon variable. We also consider
a rotation of spin operators in the $x-z$ plane at an angle
$\theta_i$ with the $z$-axis so that the classical spin vector in
the rotated frame is aligned along the $z$-axis. Under this
rotation, the components of the spin are related by,
\begin{eqnarray}
	\hat{S}_{ix}&=&  \hat{\widetilde{S}}_{ix}\cos\theta_i+
	\hat{\widetilde{S}}_{iz}\sin\theta_i\nonumber\\
	\hat{S}_{iz}&=&  \hat{\widetilde{S}}_{iz}\cos\theta_i-
	\hat{\widetilde{S}}_{ix}\sin\theta_i\nonumber\\
	\hat{S}_{iy}&=& \hat{\widetilde{S}}_{iy} \label{rot1}
\end{eqnarray}
We then use a standard Holstein-Primakoff (HP) transformation \cite{HP_transformation} of the spin operators given by,
\begin{eqnarray}
	\hat{\widetilde{S}}_{iz}&=&
	S-\hat{b}_i^{\dagger}\,\hat{b}_i\nonumber\\\
	\hat{\widetilde{S}}_{i+}&=&\sqrt{2S-\hat{b}_i^{\dagger}\,\hat{b}}_i\,\,\hat{b}_i\nonumber\\
	\hat{\widetilde{S}}_{i-}&=&\hat{b}_i^{\dagger}\sqrt{2S-\hat{b}_i^{\dagger}\,\hat{b}_i}.
	\label{hp1}
\end{eqnarray}
Finally, the Hamiltonian given in Eq.\eqref{Hamiltonian} can be
expanded in powers of $S$ leading to
\begin{eqnarray}
	\hat{\mathcal{H}} &=&
	S\,\mathcal{H}_0+\sqrt{S}\,\hat{\mathcal{H}}_1+\hat{\mathcal{H}}_2+\mathcal{O}(1/S)
	+... ,
	\label{expham1}
\end{eqnarray}
where the ellipsis represents higher powers of $1/S$. The second
term $\hat{\mathcal{H}}_1$ in the Hamiltonian proportional to
$\sqrt{S}$ vanishes for the ground state. Therefore the leading
order quantum correction around the classical ground state is,
\begin{eqnarray}
	\hat{\mathcal{H}}_{2} &=&\sum_i\left( (\omega-\mu)\,a_i^{\dagger}\,a_i+\Gamma_2\,b_i^{\dagger}\,b_i+\frac{\lambda_1}{2}(a_i\,b_i+a_i^{\dagger}\,b_i^{\dagger})\nonumber\right.\\
	&-&\left.\frac{\lambda_2}{2}(a_i\,b_i^{\dagger}+a_i^{\dagger}\,b_i)\right)-J\sum_{\langle
		ij\rangle}(a_i^{\dagger}a_j+a_j^{\dagger}a_i)\label{flucham}
\end{eqnarray}
The above Hamiltonian can be rewritten in momentum space as,
\begin{eqnarray}
	\hat{\mathcal{H}}_2 &=& \sum_{\vec{q}}\left(\Gamma_{1\vec{q}}\,a_{\vec{q}}^{\dagger}\,a_{\vec{q}}+\Gamma_2\, b_{\vec{q}}^{\dagger}\,b_{\vec{q}}\right.\nonumber\\
	&+&\left.\frac{\lambda_1}{2}(a_{\vec{q}}\,b_{-{\vec{q}}}+a_{\vec{q}}^{\dagger}\,b_{-{\vec{q}}}^{\dagger})-\frac{\lambda_2}{2}(a_{\vec{q}}\,b_{\vec{q}}^{\dagger}+a_{\vec{q}}^{\dagger}\,b_{\vec{q}})\right)\qquad
\end{eqnarray}
where $\epsilon_{\vec{q}} = 2J\sum_{j=1}^d\cos q_j$ and
$\Gamma_{1\vec{q}}=(\omega-\mu-\epsilon_{\vec{q}})$. The other
parameters have to be separately defined for different ground
states. For the ${\rm NP}_0$ phase,
\begin{eqnarray}
	\Gamma_2 &=& \omega_0-\mu, \,\,\lambda_1=0, \,\, \lambda_2=2\lambda,
	\label{np0param}
\end{eqnarray}
while for the ${\rm NP}_1$ phase
\begin{eqnarray}
	\Gamma_2 &=& -(\omega_0-\mu), \,\, \lambda_1=2\lambda, \,\,
	\lambda_2=0. \label{np1param}
\end{eqnarray}
In the superradiant phase, one has
\begin{eqnarray}
	\Gamma_2 &=& |\omega_0-\mu| \frac{\lambda^2}{\lambda_c^2}, \quad
	\lambda_{1(2)} =
	\lambda\left(1-(+)\frac{\lambda_c^2}{\lambda^2}\right).
	\label{srparam}
\end{eqnarray}
The Hamiltonian $\hat{\mathcal{H}}_2$ is quadratic in bosonic
operators. It can therefore be diagonalized and obtain the
excitation energies by the following  Bogoliubov transformation,
\begin{eqnarray}
	a_{\alpha {\vec{q}}} = \sum_{\beta}(A_{\alpha
		\beta}d_{\beta,{\vec{q}}}+\bar{A}_{\alpha
		\beta}d_{\beta,-{\vec{q}}}^{\dagger}),
\end{eqnarray}
where we denote the original bosonic operators by $a_{\alpha
	{\vec{q}}} = (a_{\vec{q}}, b_{\vec{q}} )$ and $A_{\alpha \beta}$,
$\bar{A}_{\alpha \beta}$ are elements of complex matrices ensuring
the canonical commutation relations between new bosonic operators,
$[d_{\alpha {\vec{q}}} , d_{\beta {\vec{q}}'}^{\dagger}] =
\delta_{\alpha \beta} \delta_{q,-{\vec{q}}'}$ and $[d_{\alpha
	{\vec{q}}},d_{\beta {\vec{q}}'}] = 0$. In terms of these new
operators $d_{\beta {\vec{q}}} , d_{\beta {\vec{q}}}^{\dagger}$, the
Hamiltonian $\hat{\mathcal{H}}_2$ can be written in the diagonal form as,
\begin{eqnarray}
	\hat{\mathcal{H}}_2=\sum_{\beta,
		{\vec{q}}}\mathcal{E}_{\beta}({\vec{q}})d_{\beta,{\vec{q}}}^{\dagger}d_{\beta,{\vec{q}}},
\end{eqnarray}
where the excitation energies $\mathcal{E}_{\beta}({\vec{q}})$ are
given by,
\begin{eqnarray}
	&&\mathcal{E}_{\pm}^2({\vec{q}}) =
	\frac{1}{2}\left[\Gamma_{1\vec{q}}^2+\Gamma_2^2-\frac{1}{2}(\lambda_1^2-\lambda_2^2)\right. \label{Excitation_spectrum}\\
	&\pm&\left.\sqrt{(\Gamma_{1\vec{q}}^2-\Gamma_2^2)^2-\lambda_1^2(\Gamma_{1\vec{q}}-\Gamma_2)^2+\lambda_2^2(\Gamma_{1\vec{q}}+\Gamma_2)^2}\right],\nonumber
\end{eqnarray}
which are obtained by following the standard procedure outlined in
Ref.\cite{hpref1}.

In the following subsections, we discuss the characteristic features
of the excitation spectrum of the different phases across the
critical line of phase transition. We note that the boundary between
the normal and the superradiant phase can be obtained from the
condition $\mathcal{E}_{\pm}(\vec{q}= 0)=0$. As we shall see,
similar to the scaling behavior of the Bose-Hubbard model, close to the
NP-Superradiant boundary, energy gap of the lower mode vanishes as
$\Delta_{\rm NP}\sim (\lambda_c-\lambda)^{z\nu}$, where $\nu=1/2$
for mean field result and $z$ is the dynamical critical exponent. This
scaling behavior is also reflected in the nature of the energy
dispersion at low momentum as $\mathcal{E}({\vec{q}})\sim |\vec
q|^z$.

\begin{figure}
	\includegraphics[width=\columnwidth]{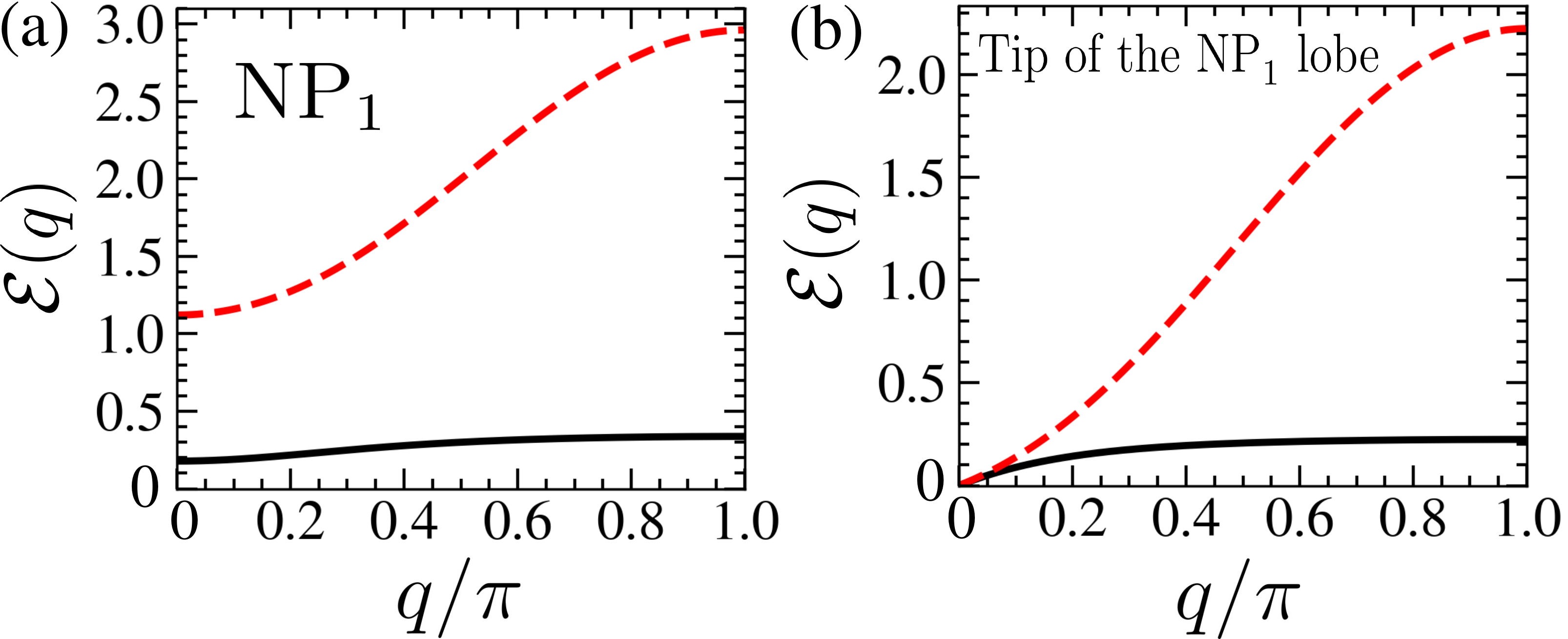}
	\caption{The low lying excitations for the normal phases as a
		function of momenta along the diagonal of the Brillouin zone
		($q_i=q$ for all $i$). (a) The spectrum of ${\rm
		NP}_0$ inside the normal phase lobe ($\mu=0.1,\lambda=0.4$). (b) Analogous plot for ${\rm NP}_1$ phase at the tip of the lobe (shown by the red dot in Fig.\ref{fig:1}).}
	\label{fig:2}
\end{figure}

\subsubsection{Normal phase}
\label{norp}

We first discuss the spectrum of the normal phases. The first of
these, ${\rm NP}_0$, occurs for $\mu\le\omega_0$ and has
$s_z^{\ast}=-1$ while the second, ${\rm NP}_1$ corresponds to
$s_z^{\ast}=1$ and occurs for
$\omega_0\le \mu\le \omega-Jz_0$. In these normal phases, we have
two gapped excitation energies (see Fig.\ \ref{fig:2}(a)) and the
dispersion of these two modes for $|\vec q| \rightarrow 0$ are given
by,
\begin{eqnarray}
	\mathcal{E}_{\pm}({\vec{q}})= \Delta_{\rm
		NP}^{\pm}(\mu,\lambda)+\tilde{c}_{\pm}\,|\vec q|^2
	\label{Dispersion_Mott}
\end{eqnarray}
where the energy gap $\Delta_{\rm NP}^{\pm}$ and the parameter
$\tilde{c}_{\pm}$ are given by,
\begin{eqnarray}
	\Delta_{\rm NP}^{\pm} &=& \frac{1}{2}\left|(\omega_0+\omega-2\mu-Jz_0\pm\sqrt{W^2-4s_z^*\lambda^2})\right|\qquad\\
	\tilde{c}_{\pm}&=&
	\frac{Jz_0}{4}\left|\left(1\pm\frac{W}{\sqrt{W^2-4s_z^*\lambda^2}}\right)\right|,
\end{eqnarray}
with $W = \omega-\omega_0-Jz_0$. The distinction between the
spectrum of the two phases comes from the different values of
$s_z^*$. The boundary $\lambda_c(\mu)$ between the NP$_0$ and
SR phase can be obtained from the condition
$\Delta_{\text{NP}}^{-}(\mu,\lambda)=0$. On the other hand, upper
and lower branches of the phase boundary between NP$_1$ and SR phases can be
obtained for $\Delta_{\rm NP}^{\pm}(\mu,\lambda)=0$, respectively.
Near the phase boundary ($\lambda\rightarrow\lambda_c$), the energy
gap of the lower mode vanishes as,
\begin{eqnarray}
	\Delta_{\text{NP}}\sim (\lambda_c-\lambda)
\end{eqnarray}
indicating the mean-field like transition with $\nu=1/2$. This follows
from the quadratic dispersion of the modes which sets dynamical
critical exponent $z=2$. The higher energy mode retains a gap
$\Delta_{\text{NP}}^{+} = |\omega+\omega_0-2\mu-Jz_0|$ at the
critical point.

However, there is a special point on the phase boundary of NP$_1$, where the critical behavior of the transition is different. There are two branches of critical lines for NP$_1$ phase,
both of which meet at the tip of the lobe given by Eq.\eqref{tip-lobe} (shown as red dot in Fig.\ref{fig:1}). Close to this point $\{\lambda^*,\mu^*\}$, both the energy gap $\Delta_{\rm
NP}^{\pm}$ and the energy dispersion ${\mathcal E}_{\vec q}$ behave as,
\begin{eqnarray}
	\Delta_{\text{NP}}(\lambda) &\sim&
	\sqrt{2\lambda^*}(\lambda^*-\lambda)^{\frac{1}{2}},
	\nonumber\\
	\mathcal{E}({\vec{q}})&\sim& \sqrt{Jd\lambda^*} |\vec q|.
	\label{z=1}
\end{eqnarray}
Thus we obtain a linearly dispersing excitation spectrum for low
momentum at the critical point (shown in Fig.\ref{fig:2}(b))
leading to $z=1$. This indicates the presence of a particle-hole
symmetry along the commensurate excitation number. Indeed, the
energy cost of creating or annihilating a photon for $\{\lambda,\mu
\}=\{\lambda^*,\mu^*\}$ can be shown to be equal; this leads to a
particle-hole symmetric dispersion which, in turn, changes the
universality class of the transition \cite{subirqpt}.

\begin{figure}
	\includegraphics[width=0.6\columnwidth]{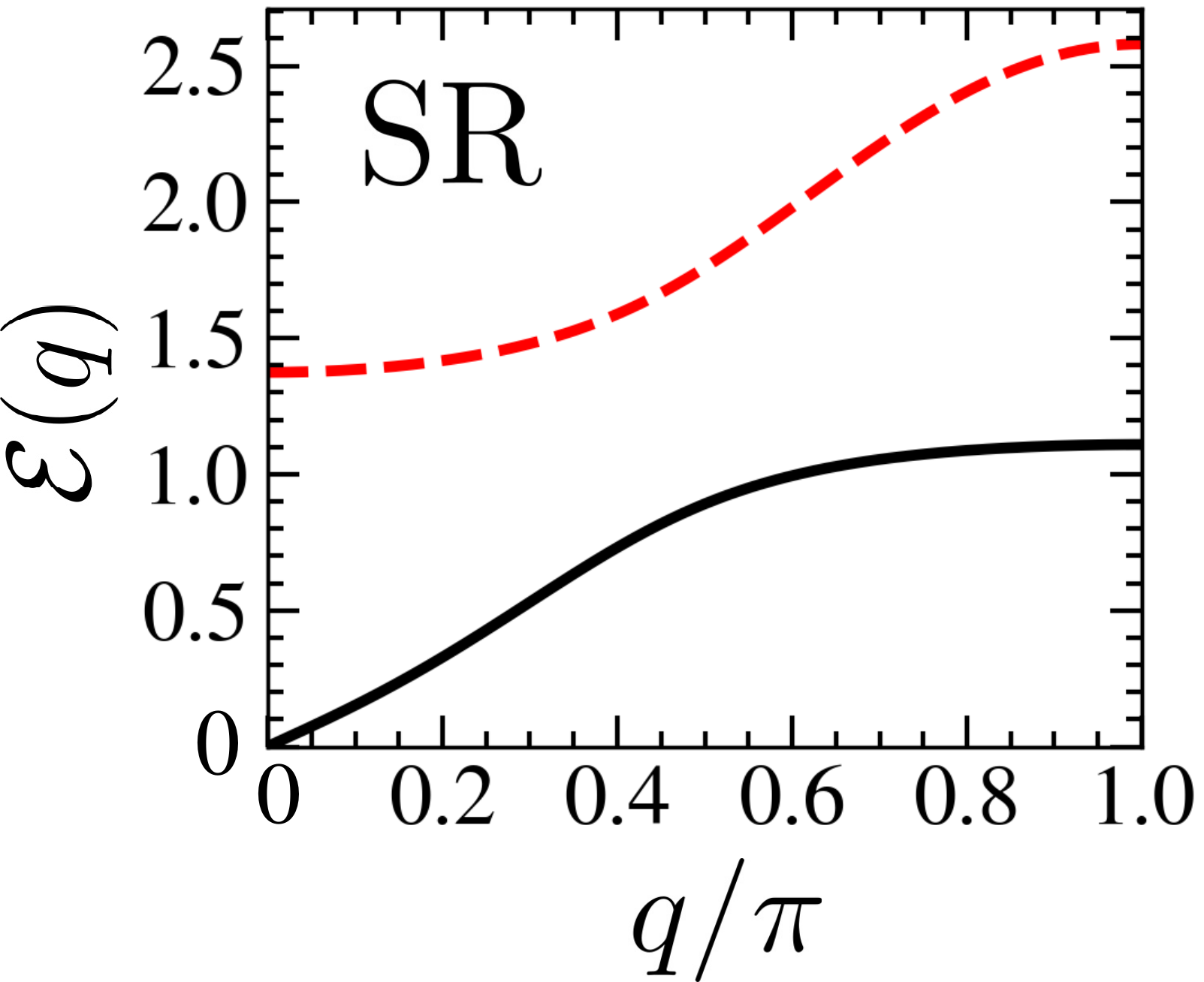}
	\caption{The low-lying collective excitations for the SR phase as a
		function of momenta along the diagonal of the Brillouin zone
		($q_i=q$ for all $i$) showing the presence of both the Higgs (red
		dashed line) and the Goldstone (black solid line) mode. For this
		plot, we have chosen $ \lambda = 0.8$ and $\mu = 0.5$.}
	\label{fig:3}
\end{figure}

\subsubsection{Superradiant phase}
\label{srp}

In contrast to the normal phases, the superradiant phase supports
both gapless Goldstone (black line in Fig.\ref{fig:3}) and massive
amplitude (Higgs) modes (red dashed line in Fig.\ref{fig:3}). This
signifies the spontaneous breaking of continuous $U(1)$ symmetry.
The Goldstone mode has a linear low energy dispersion,
\begin{eqnarray}
	\mathcal{E}_{\rm G}({\vec{q}})&=& c_s |\vec q| \label{golddis}
\end{eqnarray}
where $c_s$ is the sound velocity. Our analysis indicates that $c_s$
is given by,
\begin{eqnarray}
	c_s &=&
	\sqrt{\frac{Jz_0\tilde{\omega}(\lambda^4-\lambda_c^4)}{2(\tilde{\omega}^4+\lambda^4+2(\omega_0-\mu)\tilde{\omega}^3)}},
	\label{sound_velocity}
\end{eqnarray}
where $\tilde{\omega} = \omega-\mu-Jz_0$. It is evident from Eq.\eqref{sound_velocity} that $c_s$ vanishes at the critical point
$\lambda_c(\mu)$; consequently, the dispersion of the lower mode
becomes quadratic. The higher excitation spectrum corresponds to the
gapped Higgs mode with low energy dispersion,
\begin{eqnarray}
	\mathcal{E}_{\rm H}({\vec{q}})&=& \Delta_{\rm
		Higgs}(\mu,\lambda)+\frac{|\vec q|^2}{2m^*} \label{higgsdisp}
\end{eqnarray}
where Higgs gap $\Delta_{\rm Higgs}(\mu,\lambda)$ and the effective
mass $m^*$ are given by,
\begin{eqnarray}
	\Delta_{\rm Higgs} &=&
	\frac{1}{\tilde{\omega}}\sqrt{\tilde{\omega}^4+\lambda^4+2(\omega_0-\mu)\tilde{\omega}^3}\label{higgsparam}\\
	m^*&=&\frac{2(\tilde{\omega}^4+\lambda^4+2(\omega_0-\mu)\tilde{\omega}^3)^{3/2}}{Jz_0\tilde{\omega}^2
		(2\tilde{\omega}^4+\lambda^4+4(\omega_0-\mu)\tilde{\omega}^3+\lambda_c^4)}.
	\nonumber
\end{eqnarray}
The Higgs gap $\Delta_{\rm Higgs}$ matches with the energy gap
$\Delta^{\text{NP}}_{+}$ of the normal phase at the critical point.

Next, we focus on the excitation spectrum of the superradiant phase
along the line $\mu_c(\lambda)$ (given in Eq.\eqref{Commensurate_line}) of commensurate excitation number
$\mathcal{N}_{\rm c}=1$ (shown by black dashed line in Fig.\ref{fig:1}), which terminates at the tip of the NP$_1$ phase
boundary at $\{\lambda,\mu\}=\{\lambda^*,\mu^*\}$. The lower
excitation energy corresponds to a gapless Goldstone mode with linear
dispersion
\begin{eqnarray}
	\mathcal{E}_{\rm G}({\vec{q}})= c_s\,|\vec q|  \label{comgold1}
\end{eqnarray}
with the sound velocity,
\begin{eqnarray}
	c_s =
	\frac{3}{\sqrt{8}}\sqrt{\frac{Jz_0(\lambda^4-\lambda^{*4})\Lambda_1(\lambda)}{9\lambda^4-3\lambda^2\lambda^{*2}-2\lambda^{*3}\Lambda_1(\lambda)}}
	\label{comgold},
\end{eqnarray}
where $\Lambda_1(\lambda)=\lambda^{*}+\sqrt{\lambda^{*2}+3\lambda^2}$.
The higher energy excitation corresponds to the Higgs mode, which
acquires a relativistic dispersion (see Fig.\ref{fig:2}(b)),
\begin{eqnarray}
	\mathcal{E}_{\rm H}({\vec{q}})= \sqrt{\Delta_{\rm
			Higgs}^2+\tilde{c}^2\,|\vec q|^2} \label{comhiggs1}
\end{eqnarray}
where, the energy gap and $\tilde{c}$ are given by,
\begin{eqnarray}
	\Delta_{\rm Higgs}&=&2\sqrt{\frac{\Lambda_2^2(\lambda)-2\lambda^*\Lambda_2(\lambda)}{3}}\label{comhiggs}\\
	\tilde{c}&=&
	\sqrt{\frac{Jz_0}{9}\left(\frac{\lambda^{*2}+\lambda^*\Lambda_2(\lambda)+6\lambda^2}{\Lambda_2(\lambda)}\right)}\nonumber
\end{eqnarray}
where, $\Lambda_2(\lambda) = \sqrt{3\lambda^2+\lambda^{*2}}$.
Interestingly, near the tip of the NP$_1$ lobe
($\lambda\rightarrow\lambda^*$), the Higgs gap  becomes vanishingly
small as $\Delta_{\rm Higgs} =
2\sqrt{\lambda^*}(\lambda-\lambda^*)^{1/2}$, which reveals the
dynamical critical exponent $z=1$ for this transition, as also
obtained from the scaling of energy gap of normal phase. At the
critical point, the sound velocity is finite
$\tilde{c}=c_s=\sqrt{Jd\lambda^*}$. Thus the NP$_1$ to Superradiant
transition across the tip of the NP lobe can be captured from the
vanishing of energy gap $\Delta^{\text{NP}_1}$ as well as the Higgs
mass $\Delta_{\rm Higgs}$ at the critical point, as depicted in
Fig.\ref{fig:2}(b) and \ref{fig:3}. We find that along the line of
commensurate excitation number, the Higgs mode displays the relativistic
dispersion (Eq.\eqref{comhiggs1}) and the gapless Goldstone mode
is linearly dispersing (Eq.\eqref{comgold1}), these features distinguish
this special line from other points in the SR phase.

\section{Collective modes around the steady states in presence of dissipation}
\label{appb}

\begin{figure}[h]
	\centering
	\includegraphics[width=\columnwidth]{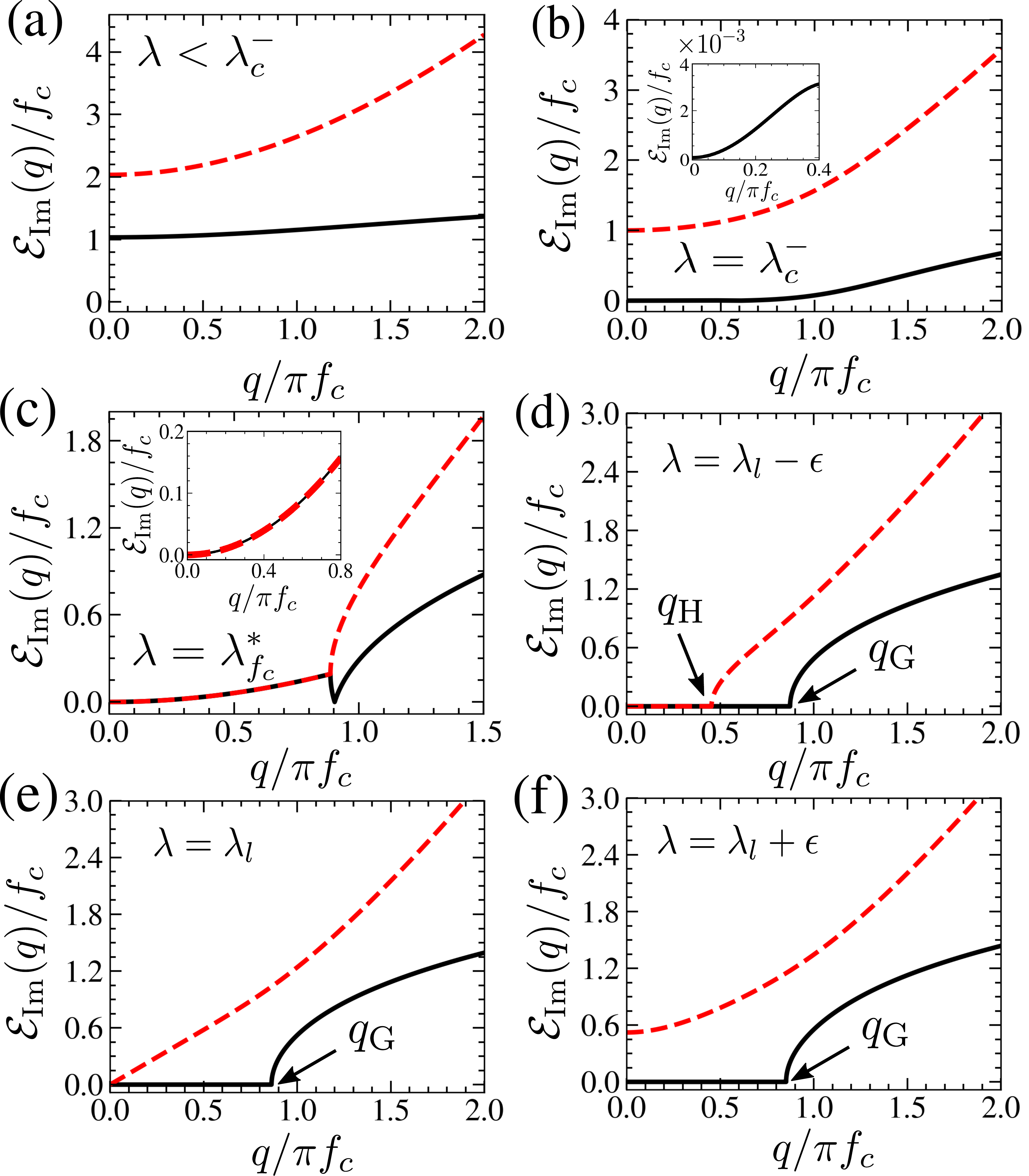}
	\caption{Plot of the excitation spectrum of normal phase NP$_1$ in
		the low momentum regime for (a) $\lambda<\lambda_c^-$ and (b)
		$\lambda=\lambda_c^-$. The inset of (b) shows the gapless quadratic
		dispersion at small $q$ on the phase boundary $\lambda_c^-(\kappa)$.
		(c-f) The low-energy dispersion of excitation spectrum of SR$_{\rm S}$
		phase along the line of commensurate excitation number in the
		presence of dissipation and pumping. The Goldstone mode is always
		overdamped away from the critical point
		($\lambda_{f_c}^*,\kappa^*$), where the imaginary part of the
		excitation energy vanishes up to a critical momentum $q_{\rm G}$
		(see main text). (c) The low-energy dispersion of the excitation
		modes at the critical point $\lambda^*_{f_c}$. The inset shows the
		quadratic nature of dispersion up to a critical momentum. (e) The
		dispersion of the higher energy mode at a special point $\lambda_l =
		\lambda_{f_c}^*+f_c^2/4\lambda^*$ away from the critical point; this
		shows a linear behavior. (d) The higher energy mode in the regime
		$\lambda<\lambda_l$ showing overdamped behavior for $q\le q_{\rm
			H}$. (f) A gap opens up at $q=0$ for $\lambda>\lambda_l$, indicating
		Higgs like behavior. Parameter chosen: $f_c=0.1$.}
	\label{fig:4}
\end{figure}

We investigate the collective excitation modes around the stable steady states for the dissipative TCH model, from the linear stability analysis, as outlined in Sec.\ref{excitation_spectrum} of the main text. First, we discuss the excitation spectrum of the normal phase NP$_1$ (see the main text for details), which is given by,
\begin{eqnarray}
	\mathcal{E}^{\pm}(q) &=& \frac{1}{4}\left[-(\kappa+2f_c)\pm 2
	i(\omega_q-2\Delta)\pm \sqrt{\mathcal{A}}\right], \nonumber\\
	\mathcal{A} &=& (2f_c-\kappa)^2+16\lambda^2-4(\omega_q+2\lambda^*)^2
	\nonumber\\
	&& \pm 4i(\omega_q+2\lambda^*)(\kappa-2f_c),
\end{eqnarray}
where $\omega_q = Jz_0(1-\cos q)$ and $\Delta =
\lambda^{\ast}(2f_c-\kappa)/(2f_c+\kappa)$. The real part of the
spectrum is negative, ensuring the stability of NP$_1$ phase. To
study the excitation spectrum, we consider a cut $q_i=q$ in the
Brillouin zone.  Similar to the non-dissipative case, there are two
gapped excitation frequencies $\mathcal{E}_{\rm Im}^{\pm}(q)$ of
NP$_1$ phase (see Fig.\ref{fig:4}(a)) and the corresponding low
energy dispersions are given by,
\begin{eqnarray}
	\mathcal{E}_{\rm Im}^{\pm}(q) = \Delta_{\rm
		NP_1}^{\pm}(\lambda,\kappa)+c_{\pm}q^2,
\end{eqnarray}
where, the energy gap $\Delta_{\rm NP_1}^{\pm}$ and $c_{\pm}$ can be
written as,
\begin{subequations}
	\begin{eqnarray}
		\Delta_{\rm NP_1}^{\pm} &=& |\Delta|\pm\frac{1}{4}\sqrt{\frac{\beta_1+\sqrt{\beta_2}}{2}-(\kappa-2f_c)^2},\\
		c_{\pm} &=& \frac{Jz_0}{4}\left(1\mp
		\frac{2\sqrt{2}\lambda^*(1+\beta_1/\sqrt{\beta_2})}{\sqrt{\beta_1+\sqrt{\beta_2}-2(\kappa-2f_c)^2}}\right),\qquad\\
		\beta_1 &=& (\kappa-2f_c)^2-16(\lambda^2-\lambda^{*2}), \\
		\beta_2 &=& \beta_1^2 +64 (\kappa-2f_c)^2 \lambda^{*2}.
	\end{eqnarray}
\end{subequations}
The lower energy mode exhibits gapless quadratic dispersion
$\mathcal{E}_{\rm Im}^{-}(q)=c_-q^2$ at the phase boundary
$\lambda_c^-(\kappa)$ between NP$_1$ and the SR$_{\rm S}$ phase (as
shown in Fig.\ref{fig:4}(b)), where,
\begin{eqnarray}
	c_- =
	Jz_0f_c\left(\frac{(\kappa+2f_c)^3+16\lambda^{*2}(2f_c-\kappa)}{
		(\kappa+2f_c)^4+16\lambda^{*2}(2f_c-\kappa)^2}\right).
\end{eqnarray}

On the other hand, the higher energy mode remains gapped as depicted
in Fig.\ref{fig:4}(b), where, the energy gap $\Delta_{\rm
	NP_1}^{+}$ and $c_+$ can be written as,
\begin{subequations}
	\begin{eqnarray}
		\Delta_{\rm NP_1}^{+} &=& 2\lambda^*\left(\frac{|2f_c-\kappa|}{2f_c+\kappa}\right)\\
		c_+ &=&
		\frac{Jz_0\kappa}{2}\left(\frac{(\kappa+2f_c)^3-16\lambda^{*2}(2f_c-\kappa)}{(\kappa+2f_c)^4+16\lambda^{*2}(2f_c-\kappa)^2}\right).\qquad
	\end{eqnarray}
\end{subequations}
Therefore, the nature of the dissipative transition between NP$_1$
and SR$_{\rm S}$ is similar to that of the non-dissipative case with
dynamical exponent $z=2$.

The characteristic features of the excitation spectrum at the critical point $(\lambda^*_{f_c},\kappa^*)$ on the boundary between the normal phase NP$_1$ and superradiant phase SR$_{\rm S}$ are discussed in Sec.\ref{excitation_spectrum} of the main text, revealing the quadratic dispersion in contrast to the non-dissipative system.
Away from the critical point along the commensurate line, both modes show overdamped dispersion with vanishing imaginary part of the spectrum up to a critical momentum $\tilde{q}_c$, which is different for the two modes. The critical momentum $\tilde{q}_c$ reduces with decreasing dissipation strengths $\kappa,f_c$ and vanishes in the non-dissipative limit. Interestingly, there is a special point $\lambda_l$ on the commensurate line, around which the Higgs mode exhibits relativistic dispersion. The approximate value of $\lambda_l$ considering small $f_c$ is given by,
\begin{eqnarray}
	\lambda_l = \lambda_{f_c}^*+\frac{f_c^2}{4\lambda^*}.
\end{eqnarray}
The energy dispersion of the Higgs and Goldstone mode close to this
point ($ \lambda-\lambda_l=\epsilon\rightarrow 0$) are given by,
\begin{eqnarray}
\mathcal{E}_{\rm H (G)}(q) = \sqrt{\lambda^*\omega_q+m_{\rm H(G)}^2
}, \label{Excitation_special_point}
\end{eqnarray}
where, the mass corresponding to the Higgs and Goldstone mode are
given by,
\begin{eqnarray}
m_{\rm H}^2 = 4\lambda^*\epsilon,\quad m_{\rm G}^2 = -f_c^2.
\end{eqnarray}
Interestingly, as a result of dissipation, the Goldstone mode
acquires an imaginary mass, due to which it is overdamped up to a
critical momentum $q_{\rm G} = f_c\sqrt{2/Jz_0\lambda^*}$ (see Fig.\ref{fig:4}(d,e,f). Note that, the imaginary part of the mode
frequencies do not affect the stability of this phase, instead it
contributes to $\mathcal{E}_{\rm Re}$ and stability is ensured by
the condition $\mathcal{E}_{\rm Re}<0$. For $\lambda<\lambda_l
(\epsilon<0)$, the Higgs mode has also become overdamped up to a
critical value of momentum $q_{\rm H} = \sqrt{8|\epsilon|/Jz_0}$, as
shown in Fig.\ref{fig:4}(d).
At $\lambda = \lambda_l$, the Higgs mode acquires a linear
dispersion (see Fig.\ref{fig:4}(e)), with the $f_c$ independent term of the sound velocity $c_s = \sqrt{Jz_0\lambda^*/2}$, same as that of the non-dissipative case at
the critical point $\lambda^*$. For $\lambda>\lambda_l$, a gap
$m_{\rm H}\sim \sqrt{\lambda-\lambda_l}$ opens up for the Higgs
mode; in contrast, the Goldstone mode remains overdamped up to the
momentum $q_{\rm G}$, which is depicted in Fig.\ref{fig:4}(f). The
above features of the collective modes can serve as a signature to
identify dissipative phases and the dynamical transition between
them.

\end{document}